\documentstyle[12pt]{article}
\def\lsim{\mathrel{\rlap {\raise.5ex\hbox{$ < $}}
{\lower.5ex\hbox{$\sim$}}}}

\newcommand{\be}{\begin{equation}}
\newcommand{\ee}{\end{equation}}
\newcommand{\bea}{\begin{eqnarray}}

\newcommand{\eea}{\end{eqnarray}}

\def\gappeq{\mathrel{\rlap {\raise.5ex\hbox{$>$}}
{\lower.5ex\hbox{$\sim$}}}}

\def\lappeq{\mathrel{\rlap{\raise.5ex\hbox{$<$}}
{\lower.5ex\hbox{$\sim$}}}}

\def\ie{{\em i.e.}}
\def\eg{{\em e.g.}}
\def\s{{\,\rm s}}
\def\S{S\hskip-8pt/\hskip2pt}
\def\H{H\hskip-8.5pt/\hskip2pt}

\def\beq{\begin{equation}}
\def\eeq{\end{equation}}

\def\I#1{{\rm Im}\,#1}
\def\Tr{{\rm Tr}\,}
\catcode`\@=11 
\def\coeff#1#2{{\textstyle{#1\over #2}}}

\def\ket#1{\left| #1\right\rangle}
\def\VEV#1{\left\langle #1\right\rangle}
\def\vev#1{\left\langle #1\right\rangle}
\def\lsim{\mathrel{\mathpalette\@versim<}}
\def\gsim{\mathrel{\mathpalette\@versim>}}
\def\@versim#1#2{\vcenter{\offinterlineskip
    \ialign{$\m@th#1\hfil##\hfil$\crcr#2\crcr\sim\crcr } }}
\def\etal{{\em et. al.}}
\def\JL{J. L. Lopez}
\def\DVN{D. V. Nanopoulos}

\def\t1{{\tilde 1}}

\def\GeV{\,{\rm GeV}}

\def\to{\rightarrow}

\def\NPB#1#2#3{Nucl. Phys. B {\bf#1} (19#2) #3}
\def\PLB#1#2#3{Phys. Lett. B {\bf#1} (19#2) #3}

\def\PRD#1#2#3{Phys. Rev. D {\bf#1} (19#2) #3}
\def\PRL#1#2#3{Phys. Rev. Lett. {\bf#1} (19#2) #3}

\def\gappeq{\mathrel{\rlap {\raise.5ex\hbox{$>$}}
{\lower.5ex\hbox{$\sim$}}}}

\def\lappeq{\mathrel{\rlap{\raise.5ex\hbox{$<$}}
{\lower.5ex\hbox{$\sim$}}}}

\begin{document}

\begin{flushright}
\baselineskip=12pt
ACT-06/95 \\
CERN-TH.95-99 \\
CTP-TAMU-16/95 \\
ENSLAPP-A-521/95 \\
{\tt hep-ph/9505340}
\end{flushright}

\begin{center}
\vglue 0.25cm
{\Large\bf Precision Tests of CPT Symmetry and Quantum Mechanics in the Neutral
Kaon System \\}
\vglue 0.5cm
{\large John Ellis$^{(a)}$, Jorge L. Lopez$^{(b,c)}$, N. E.
 Mavromatos$^{(d,e)}$, and D.~V.~Nanopoulos$^{(b,c)}$\\}
\vglue 0.2cm
\end{center}

\begin{flushleft}
\small
{$^{(a)}$ CERN Theory Division, 1211 Geneva 23, Switzerland\\}
{$^{(b)}$ Center for Theoretical Physics, Department of Physics, Texas
A\&M University\\}
{College Station, TX 77843--4242, USA\\}
{$^{(c)}$ Astroparticle Physics Group, Houston Advanced Research Center
(HARC)\\}
{The Mitchell Campus, The Woodlands, TX 77381, USA\\}
{$^{(d)}$ Laboratoire de Physique Th\'eorique
ENSLAPP (URA 14-36 du CNRS, associ\'ee \`a l' E.N.S
de Lyon, et au LAPP (IN2P3-CNRS) d'Annecy-le-Vieux),
Chemin de Bellevue, BP 110, F-74941 Annecy-le-Vieux
Cedex, France\\}
{$^{(e)}$ On leave from P.P.A.R.C. Advanced Fellowship, Dept. of Physics
(Theoretical Physics), University of Oxford, 1 Keble Road,
Oxford OX1 3NP, U.K.  \\}
\end{flushleft}
\vspace{0.5cm}
\begin{abstract}
We present a systematic phenomenological analysis of the tests of CPT symmetry
that are possible within an {\em open} quantum-mechanical description of the
neutral kaon system that is motivated by arguments based on quantum gravity and
string theory. We develop a perturbative expansion in terms of the three
small CPT-violating parameters admitted in this description, and provide
expressions for a complete set of $K \rightarrow 2\pi, 3\pi$ and $\pi\ell\nu$
decay observables to second order in these small parameters. We also illustrate
the new tests of CPT symmetry and quantum mechanics that are possible in this
formalism using a regenerator. Indications are that experimental data from the
CPLEAR and previous experiments could be used to establish upper bounds on the
CPT-violating parameters that are of order $10^{-19}$ GeV, approaching the
order of magnitude that may be attainable in quantum theories of gravity.
\end{abstract}
\vspace{0.5cm}

\begin{flushleft}
\baselineskip=12pt
ACT-06/95 \\
CERN-TH.95-99 \\
CTP-TAMU-16/95 \\
ENSLAPP-A-521/95 \\
May 1995
\end{flushleft}

\vfill\eject
\setcounter{page}{1}
\pagestyle{plain}
\baselineskip=14pt

\section{Introduction}
\label{sec:intro}
The neutral kaon system has long served as a penetrating probe of
fundamental physics. It has revealed or illuminated many new areas of
fundamental physics, including parity violation, CP violation, flavour-changing
neutral interactions, and charm. It remains the most sensitive test of
fundamental symmetries, being the only place where CP violation has been
observed, namely at the level of $10^{-18}$ GeV in the imaginary part of the
effective mass matrix for neutral kaons, and
providing the most stringent microscopic check
of CPT symmetry within the framework of quantum mechanics,
namely $|(m_{K^0}-m_{\bar K^0})/m_{K^0}|<9\times10^{-19}$
\cite{pdg}.

It is well known that CPT symmetry is a fundamental theorem of quantum field
theory, which follows from locality, unitarity, and Lorentz invariance
\cite{lud}. However, the topic of CPT violation has recently attracted
increased attention,
drawn in part by the prospect of higher-precision tests by
CPLEAR~\cite{Guyot} and at DA$\Phi$NE~\cite{dafnehb},
and in part by the renewed theoretical interest in quantum gravity motivated by
recent developments in string theory. Some of the phenomenological discussion
has been in the context of quantum mechanics \cite{peccei}, abandoning
implicitly or explicitly the derivation of quantum mechanics from quantum field
theory, in which CPT is sacrosanct. Instead, we have followed the approach of
Ref.~\cite{EHNS}, in which a parametrization of CPT-violating effects is
introduced via a deviation from conventional quantum mechanics
\cite{EHNS,emnqm} believed to reflect the loss of quantum coherence expected in
some approaches to quantum gravity \cite{Hawking}, notably one based on
a non-critical formulation of string theory \cite{emnerice}.

The suggestion that quantum coherence might be lost at the microscopic
level was made in Ref.~\cite{Hawking}, which suggested that asymptotic
scattering should be described in terms of a superscattering operator $\S$,
relating initial $(\rho_{\rm in})$ and final $(\rho_{\rm out})$ density
matrices, that does not factorize as a product of $S$- and $S^\dagger$-matrix
elements:
\begin{equation}
\rho_{\rm out} = \S \rho_{\rm in} :\quad \S \not= SS^\dagger\ .
\label{1.1}
\end{equation}
The loss of quantum coherence was thought to be a consequence of
microscopic quantum-gravitational fluctuations in the space-time background.
Model calculations supporting this suggestion were presented~\cite{Hawking} and
contested~\cite{Gross}. Ref.~\cite{EHNS}
pointed out that if Eq.~(\ref{1.1}) is correct for asymptotic
scattering, there should be a corresponding effect in the quantum Liouville
equation that describes the time-evolution of the dentity matrix $\rho (t)$:
\begin{equation}
{\partial\rho(t)\over\partial t}=i[\rho,H]+i{\delta\H}\rho\ ,
\label{1.2}
\end{equation}
which is characteristic of an open quantum-mechanical system. Ref.~\cite{EHNS}
parametrized the non-Hamiltonian term in the case of a simple
two-state system such as the $K^0-\bar K^0$ system, presented a first analysis
of its phenomenological consequences, and gave experimental bounds on the
non-quantum-mechanical parameters.

The question of microscopic quantum coherence
has recently been addressed in the context
of string theory using a variety of approaches \cite{cptstrings}.
In particular, we have analyzed this question using non-critical
string theory \cite{aben}, with criticality restored by non-trivial
dynamics for a time-like Liouville field \cite{aben,DDK}, which we identify
with the world-sheet cutoff and the target time variable \cite{emnqm,emnerice}.
This approach leads to an equation of the form (\ref{1.2}), in which
probability and energy are conserved, and the possible magnitude of
the extra term $|\delta\H|={\cal O}(E^2/M_{Pl})$, where $E$ is a
typical energy scale of the system under discussion. The details
of this approach are not essential for the phenomenological
discussion of this paper, but it is interesting to note that the experimental
sensitivity may approach this theoretical magnitude.

It has been pointed out \cite{wald} that at least the strong version
of the CPT theorem must be violated in any theory described by a
non-factorizing superscattering matrix $\S$ (\ref{1.1}), which leads to
a loss of quantum coherence. This is also true of the parametrization proposed
by Ref.~\cite{EHNS}, which violates CPT in an
intrinsically non-quantum-mechanical way. More detailed descriptions of
phenomenological implications and improved experimental bounds were
presented in Ref.~\cite{emncpt}. These results were based on an analysis of
$K_L$ and $K_S$ decays, and did not consider the additional constraints
obtainable from an analysis of intermediate-time data. A systematic approach to
the time evolution of the density matrix for the neutral kaon system was
proposed in Ref.~\cite{Lopez}, and preliminary estimates of the improved
experimental constraints on the non-quantum-mechanical parameters were
presented. Similar results were presented later in Ref.~\cite{HP}, which also
discussed correlation measurements possible at a $\phi$ factory such as
DA$\Phi$NE.

The main focus of this paper is to present detailed formulae for the
time dependences of several decay asymmetries that can be measured by the
CPLEAR and DA$\Phi$NE experiments, using the systematic approach proposed in
Ref.~\cite{Lopez} and described in Section~\ref{sec:perturbation}. In
particular, we discuss in Section~\ref{sec:analytical} the asymmetries known as
$A_{2\pi}, A_{3\pi}, A_{\rm T}, A_{\rm CPT}$ and $A_{\Delta m}$, whose
definitions are reviewed in Section~\ref{sec:observables}. We show in
Section~\ref{sec:regeneration} that experiments with a regenerator can provide
useful new measurements of the non-quantum-mechanical CPT-violating parameters.
Then, in Section~\ref{sec:Bounds} we derive illustrative bounds on the
non-quantum-mechanical parameters from all presently available data.
Section~\ref{sec:HP} contains a brief discussion of the extension of the
formalism of Ref.~\cite{EHNS} to the correlation measurements possible at
$\phi$ factories such as DA$\Phi$NE. We emphasize the need to consider a
general parametrization of the two-particle density matrix, that cannot be
expressed simply in terms of the previously-introduced single-particle density
matrix parameters, and enables energy conservation to be maintained, as we have
demonstrated \cite{emnqm,emnerice} in our non-critical string theory approach
to the loss of quantum coherence. In Section~\ref{sec:conclusions} we
review our conclusions and discuss the prospects for future experimental and
theoretical work. Formulae for the CPLEAR observables in the context of
standard quantum-mechanical CPT violation \cite{peccei} are collected in
Appendix~\ref{app:A}, where bounds on the corresponding parameters are also
obtained. Lastly, complete formulae for the second-order contributions to the
density matrix in our quantum-mechanical-violating framework are collected in
Appendix~\ref{app:B}.

\section{Formalism and Relevant Observables}
\label{sec:observables}
In this section we first
review aspects of the modifications (\ref{1.2}) of
quantum mechanics believed to be induced by quantum
gravity \cite{EHNS}, as argued
specifically in the
context of a non-critical string analysis \cite{emnqm,emnerice}.
This provides a specific
form for the modification (\ref{1.2}) of the quantum
Liouville equation for the temporal evolution of the density matrix of
observable matter \cite{emnqm,emnerice}
\begin{equation}
   \frac{\partial }{\partial t} \rho = i [ \rho, H] + i{\delta\H} \rho
\qquad ; \qquad \delta\H \equiv {\dot g}^i G_{ij} [g^i, \rho ]
\label{one}
\end{equation}
where the coordinates $\{ g^i\} $
parametrize the space of possible
string models and the
extra term $\delta\H$ is such that
the time evolution has
the following basic properties:
\begin{description}
\item (i) The total probability is {\it conserved} in time
\begin{equation}
       \frac{\partial }{\partial t} \Tr \rho = 0
\label{2.1}
\end{equation}
\item (ii) The energy $E$ is {\it conserved on the average}
\begin{equation}
  \frac{\partial }{\partial t} \Tr(E \rho ) = 0
\label{2.2}
\end{equation}
as a result of
the {\it renormalizability} of the world-sheet $\sigma$-model
specified by the parameters $g^i$ which
describe string propagation in a string space-time foam background.
\item (iii) The von Neumann entropy $S \equiv -k_B \Tr \rho \ln\rho $
increases {\it monotonically with time}
\begin{equation}
    \frac{\partial }{\partial t} S \ge 0
\label{2.3}
\end{equation}
which vanishes only if one restricts one's attention to critical (conformal)
strings, in which case there is no arrow of time \cite{emnqm,emnerice}.
However, we argue that quantum fluctuations in the background space time should
be treated by including non-critical (Liouville) strings \cite{aben,DDK}, in
which case (\ref{2.3}) becomes a strict inequality. This latter property also
implies that the statistical entropy $S_{\rm st} \equiv \Tr \rho ^2 $ is also
monotonically increasing with time, pure states evolve into mixed ones and
there is an arrow of time in this picture \cite{emnqm}.
\item (iv) Correspondingly, the superscattering matrix $\S$, which is
defined by its action on asymptotic density matrices
\begin{equation}
  \rho _{out} = \S \rho _{in}
\label{2.4}
\end{equation}
cannot be factorised into the usual product of the Heisenberg scattering matrix
and its hermitian conjugate
\begin{equation}
     \S \ne S S^\dagger \qquad ; \qquad S=e^{-iHt}
\label{2.5}
\end{equation}
with $H$ the Hamiltonian operator of the system. In particular this property
implies that $\S$ has no inverse, which is also expected from the property
(iii).
\item (v) The absence of an inverse for $\S$ implies that {\it strong} CPT
invariance of the low-energy subsystem is lost, according
to the general analysis of \cite{wald,emnerice}.
\end{description}

It should be stressed that, although for the purposes of the present work we
keep the microscopic origin of the quantum-mechanics-violating terms
unspecified, it is only in the non-critical string model of Ref.~\cite{emnqm} -
and the associated approach to the nature of time - that a concrete microscopic
 model guaranteeing the properties (i)-(v) has so far emerged naturally.
Within this framework, we expect
that the string $\sigma$-model coordinates $g^i$ obey
renormalization-group equations of the general form
\begin{equation}
    {\dot g}^i = \beta ^i M_{Pl} \qquad : \qquad
    |\beta ^i | = {\cal O}\left(\frac{E^2}{M_{Pl}^2}\right)
\label{E}
\end{equation}
where the dot denotes differentiation
with respect to the target time, measured in string $(M_{Pl}^{-1})$
units, and
$E$ is a typical energy scale in the observable
matter system. Since $G_{ij}$ and
$g^i$ are themselves dimensionless numbers of order unity,
we expect that
\begin{equation}
      |\delta\H | = {\cal O}\left(\frac{E^2}{M_{Pl}}\right)
\label{F}
\end{equation}
in general. However, it should be emphasized that there
are expected to be system-dependent numerical factors
that depend on the underlying string model, and that
$|\delta\H |$ might be suppressed by further
($E/M_{Pl}$)-dependent factors, or even vanish.
Nevertheless, (\ref{F}) gives us an order of magnitude
to aim for in the neutral kaon system,
namely ${\cal O}((\Lambda_{\rm QCD}~{\rm or}~m_s)^2/M_{Pl}) \sim 10^{-19}$ GeV.

In the formalism of Ref.~\cite{EHNS}, the extra (non-Hamiltonian) term in the
Liouville equation for $\rho$ can be parametrized by a $4\times 4$ matrix
$\delta\H_{\alpha\beta} $, where the indices $\alpha, \beta, \dots$ enumerate
the Hermitian $\sigma$-matrices $\sigma _{0,1,2,3}$, which we represent in the
$K_{1,2}$ basis. We refer the reader to the literature \cite{EHNS,emncpt}
and Appendix~\ref{app:A} for details of this description, noting here the
following forms for the neutral kaon Hamiltonian
\begin{equation}
  H = \left( \begin{array}{cc}
  M - \coeff{i}{2}\Gamma - {\rm Re} M_{12} + \coeff{i}{2} {\rm Re} \Gamma _{12}
&  \coeff{1}{2}\delta M - \coeff{i}{4} \delta \Gamma
  -i {\rm Im} M_{12}  - \coeff{1}{2} {\rm Im} \Gamma _{12}  \\
  \coeff{1}{2}\delta M - \coeff{i}{4} \delta \Gamma
  + i {\rm Im} M_{12}  - \coeff{1}{2} {\rm Im} \Gamma _{12} &
  M - \coeff{i}{2}\Gamma + {\rm Re} M_{12} - \coeff{i}{2} {\rm Re} \Gamma _{12}
 \end{array}\right)
\label{nkham}
\end{equation}
in the $K_{1,2}$ basis, or
\begin{equation}
 H_{\alpha\beta}
 =\left( \begin{array}{cccc}  - \Gamma & -\coeff{1}{2}\delta \Gamma
& -{\rm Im} \Gamma _{12} & -{\rm Re}\Gamma _{12} \\
 - \coeff{1}{2}\delta \Gamma
  & -\Gamma & - 2{\rm Re}M_{12}&  -2{\rm Im} M_{12} \\
 - {\rm Im} \Gamma_{12} &  2{\rm Re}M_{12} & -\Gamma & -\delta M    \\
 -{\rm Re}\Gamma _{12} & -2{\rm Im} M_{12} & \delta M   & -\Gamma
\end{array}\right)
\label{hnk12}
\end{equation}
in the $\sigma$-matrix basis. As discussed in Ref.~\cite{EHNS}, we
assume that the dominant violations of quantum mechanics conserve strangeness,
so that $\delta\H_{1\beta }$ = 0, and that $\delta\H_{0\beta }$ = 0 so as to
conserve probability. Since $\delta\H_{\alpha\beta }$ is a symmetric
matrix, it follows that also $\delta\H_{\alpha 0}=\delta\H_{\alpha 1}=0$.
Thus, we arrive at the general parametrization
 \begin{equation}
  {\delta\H}_{\alpha\beta} =\left( \begin{array}{cccc}
 0  &  0 & 0 & 0 \\
 0  &  0 & 0 & 0 \\
 0  &  0 & -2\alpha  & -2\beta \\
 0  &  0 & -2\beta & -2\gamma \end{array}\right)
\label{nine}
\end{equation}
where, as a result of the positivity of the hermitian density matrix $\rho$
\cite{EHNS}
\begin{equation}
\alpha, \gamma  > 0,\qquad \alpha\gamma>\beta^2\ .
\label{positivity}
\end{equation}

We recall \cite{emncpt} that the CPT transformation can be expressed as
a linear combination of $ \sigma _{2,3}$ in the $K_{1,2}$ basis :
${\rm CPT} = \sigma_3 \cos\theta + \sigma_2 \sin\theta$,
for some choice of phase $\theta$. It is
apparent that none of the non-zero terms $\propto   \alpha ,  \beta ,
 \gamma $ in $\delta\H_{\alpha\beta}$  (\ref{nine})
commutes with the CPT transformation. In other words, each of the three
parameters $\alpha$, $\beta$, $\gamma$ violates CPT, leading to a
richer phenomenology than in conventional quantum mechanics. This is
because the symmetric $\delta\H$ matrix has three parameters in its
bottom right-hand $2\times 2$ submatrix, whereas the $h$ matrix
appearing in the time evolution within quantum mechanics \cite{peccei}
has only one complex CPT-violating parameter $\delta$,
\begin{equation}
\delta = -\coeff{1}{2}
\frac{\coeff{1}{2}\delta\Gamma+i\delta M}{\coeff{1}{2}|\Delta\Gamma|+i\Delta m}
\ ,
\label{cptdelta}
\end{equation}
where $\delta M$ and $\delta \Gamma $ violate CPT, but do not induce any mixing
in the time evolution of pure state vectors\cite{emncpt}. The parameters
$\Delta m = M_L -M_S$ and $|\Delta\Gamma|=\Gamma_S-\Gamma_L$ are the
usual differences between mass and decay widths, respectively, of $K_L$ and
$K_S$ states. A brief review of the quantum-mechanical formalism is given in
Appendix~\ref{app:A}. For more details we refer the reader to the literature
\cite{emncpt}. The above results imply that the experimental constraints
\cite{pdg} on CPT violation have to be rethought. As we shall discuss later on,
there are essential differences between quantum-mechanical CPT violation and
the non-quantum-mechanical CPT violation induced by the effective parameters
$\alpha, \beta, \gamma$ \cite{EHNS}.

Useful observables are associated with the decays of neutral kaons to $2\pi$ or
$3\pi$ final states, or semileptonic decays to $\pi l \nu$. In the
density-matrix formalism introduced above, their values are given by
 expressions of the form \cite{EHNS}
\begin{equation}
     \VEV{O_i}= {\rm Tr}\,[O_i\rho]\ ,
\label{13and1/2}
\end{equation}
where the observables $O_i$ are represented by $2 \times 2$ hermitian
matrices.
For future use, we give their
expressions in the $K_{1,2}$ basis
\begin{eqnarray}
 O_{2\pi} &=& \left( \begin{array}{cc} 0 & 0 \\
0 & 1 \end{array} \right)\ ,\qquad  O_{3\pi} \propto
\left( \begin{array}{cc} 1 & 0 \\
0 & 0 \end{array} \right)\ , \label{2pi-obs} \\
O_{\pi^-l^+\nu} &=& \left( \begin{array}{cc}
1 & 1 \\1 & 1 \end{array} \right)\ ,\qquad
O_{\pi^+l^- \bar\nu} = \left( \begin{array}{rr}
1 & -1 \\
-1 & 1 \end{array} \right)\ .
\label{semi-obs}
\end{eqnarray}
which constitute a complete hermitian set. As we discuss in more detail later,
it is possible to measure the interference between $K_{1,2}$ decays into $\pi
^{+}\pi^{-}\pi^0$ final states with different CP properties, by restricting
one's attention to part of the phase space $\Omega$, \eg, final states with
$m(\pi^+\pi^0)>m(\pi^-\pi^0)$. In order to separate this interference from that
due to $K_{S,L}$ decays into final states with identical CP properties,
due to CP violation in the $K_{1,2}$ mass matrix or in decay amplitudes, we
consider \cite{nakada} the difference between final states with
$m(\pi^+\pi^0) > m(\pi^-\pi^0) $ and $m(\pi^+\pi^0) < m(\pi^-\pi^0) $.
This observable is represented by the matrix
\begin{equation}
 O_{3\pi}^{\rm int} = \left( \begin{array}{cc} 0 & {\cal K} \\
{\cal K}^* & 0 \end{array} \right)
\label{XI}
\end{equation}
where
\begin{equation}
{\cal K} \equiv
\frac{\left[\int _{m(\pi^+\pi^0) > m(\pi^-\pi^0)} d\Omega
-\int _{m(\pi^+\pi^0) < m(\pi^-\pi^0)} d\Omega\right]
A_2 (I_{3\pi}=2)A_1 (I_{3\pi}=1)}
{\int d\Omega |A_1 (I_{3\pi}=1)|^2 }
\label{Y}
\end{equation}
where ${\cal K}$ is expected to be essentially real, so that the $O_{3\pi}^{\rm
int}$ observable provides essentially the same information
as $O_{\pi^-l^+\nu}-O_{\pi^+l^-{\overline \nu}}$.

In this formalism, pure $K^0$ or ${\bar K}^0$ states, such as the ones used as
initial conditions in the CPLEAR experiment \cite{Guyot},
are described by the following density matrices
\begin{equation}
\rho _{K^0} =\coeff{1}{2}\left( \begin{array}{cc}
1 &1 \\1 & 1 \end{array} \right)\ , \qquad
\rho _{{\bar K}^0} =\coeff{1}{2}\left( \begin{array}{rr}
1 & -1 \\-1 & 1 \end{array} \right)\ .
\label{rhos}
\end{equation}
We note the similarity of the above density matrices (\ref{rhos})
to the semileptonic decay observables in (\ref{semi-obs}), which is
due to the strange quark ($s$) content of the kaon $K^0 \ni {\bar s}
\rightarrow {\bar u} l^+ {\nu} , {\bar K}^0 \ni s \rightarrow  u l^- \bar\nu$,
and our assumption of the validity of the $\Delta S = \Delta Q$ rule.

In this paper we shall apply the above formalism to compute the time evolution
of certain experimentally-observed quantities that are of relevance to the
CPLEAR experiment \cite{Guyot}. These are asymmetries associated with
decays of an initial $K^0$ beam as compared to corresponding decays of an
initial ${\bar K}^0$ beam
\begin{equation}
    A (t) = \frac{
    R({\bar K}^0_{t=0} \rightarrow
{\bar f} ) -
    R(K^0_{t=0} \rightarrow
f ) }
{ R({\bar K}^0_{t=0} \rightarrow
{\bar f} ) +
    R(K^0_{t=0} \rightarrow
f ) }\ ,
\label{asym}
\end{equation}
where $R(K^0\rightarrow f)\equiv \Tr[O_{f}\rho (t)]$, denotes the decay rate
into the final state $f$, given that one starts from a pure $ K^0$ at $t=0$,
whose density matrix is given in (\ref{rhos}), and
$R({\bar K}^0 \rightarrow {\bar f}) \equiv \Tr [O_{\bar f} {\bar \rho}(t)]$
denotes the decay rate into the conjugate state ${\bar f}$, given that one
starts from a pure ${\bar K}^0$ at $t=0$.

Let us illustrate the above formalism by two examples. We may compute the
asymmetry  for the case where there are identical final states
$f={\bar f} = 2\pi $, in which case the observable is given in (\ref{2pi-obs}).
We obtain
\begin{equation}
A_{2\pi} = \frac{
\Tr[O_{2\pi} {\bar \rho} (t)] -
\Tr[O_{2\pi} \rho (t)]}
{
\Tr[O_{2\pi} {\bar \rho} (t)] +
\Tr[O_{2\pi} \rho (t)]}
= \frac{\Tr[O_{2\pi} \Delta \rho (t)]}{\Tr[O_{2\pi} \Sigma \rho (t)]}\ ,
\label{a2pi}
\end{equation}
where we have defined: $\Delta\rho(t)\equiv\bar\rho(t)-\rho(t)$ and
$\Sigma\rho(t)\equiv\bar\rho(t)+\rho(t)$. We note that in the above formalism
we make no distinction between neutral and charged two-pion final states.
This is because we neglect, for simplicity, the effects of $\epsilon'$.
Since $|\epsilon'/\epsilon|\lsim 10^{-3}$, this implies that our analysis
of the new quantum-mechanics-violating parameters must be refined
if magnitudes $\lsim\epsilon'|\Delta\Gamma| \simeq
10^{-6}|\Delta\Gamma|$ are to be studied.

In a similar spirit to the identical final state case, one can compute the
asymmetry $A_{\rm T}$ for the semileptonic decay case, where
$f=\pi^+l^-\bar\nu\ \not=\ \bar f=\pi^-l^+\nu$.
The formula for this observable is
\begin{equation}
A_{\rm T}(t)
={\Tr[O_{\pi^-l^+\nu}\bar\rho(t)]-\Tr[O_{\pi^+l^-\bar\nu}\rho(t)]
\over\Tr[O_{\pi^-l^+\nu}\bar\rho(t)]+\Tr[O_{\pi^+l^-\bar\nu}\rho(t)]}\ .
\label{asymt}
\end{equation}
Other observables are discussed in Section~\ref{sec:analytical}.

To determine the temporal evolution of the above observables, which is
crucial for experimental fits, it is necessary to know the equations of motion
for the components of $\rho$ in the $K_{1,2}$ basis. These are
\cite{EHNS,emncpt}\footnote{Since we neglect $\epsilon'$ effects and assume the
validity of the $\Delta S=\Delta Q$ rule, in what follows we also consistently
neglect ${\rm Im}\,\Gamma_{12}$ \cite{dafnehb}.}
\begin{eqnarray}
\dot\rho_{11}&=&-\Gamma_L\rho_{11}+\gamma\rho_{22}
-2{\rm Re}\,[({\rm Im}M_{12}-i\beta)\rho_{12}]\,,\label{rho11}\\
\dot\rho_{12}&=&-(\Gamma+i\Delta m)\rho_{12}
-2i\alpha\I{\rho_{12}}+({\rm Im}M_{12}-i\beta)(\rho_{11}-\rho_{22})\,,
\label{rho12}\\
\dot\rho_{22}&=&-\Gamma_S\rho_{22}+\gamma\rho_{11}
+2{\rm Re}\,[({\rm Im}M_{12}-i\beta)\rho_{12}]\,,   \label{rho22}
\end{eqnarray}
where for instance $\rho$ may represent $\Delta\rho$ or $\Sigma\rho$, defined
by the initial conditions
\begin{equation}
\Delta\rho(0)=\left(\begin{array}{cc}0&-1\\-1&0\end{array}\right)\ ,\qquad
\Sigma\rho(0)=\left(\begin{array}{cc}1&0\\0&1\end{array}\right)\ .
\label{InitialConditions}
\end{equation}
In these equations $\Gamma_L=(5.17\times10^{-8}\s)^{-1}$ and
$\Gamma_S=(0.8922\times10^{-10}\s)^{-1}$ are the inverse $K_L$ and $K_S$
lifetimes, $\Gamma\equiv(\Gamma_S+\Gamma_L)/2$, $|\Delta\Gamma| \equiv
\Gamma_S-\Gamma_L =(7.364 \pm 0.016 )\times 10^{-15}\GeV$, and $\Delta
m=0.5351\times10^{10}\s^{-1}=3.522\times10^{-15}\GeV$ is the $K_L-K_S$
mass difference. Also, the CP impurity parameter $\epsilon$ is given by
\begin{equation}
  \epsilon =\frac{{\rm Im} M_{12}}{\coeff{1}{2}|\Delta\Gamma|+i\Delta m }\ ,
\label{epstext}
\end{equation}
which leads to the relations
\begin{equation}
\I{M_{12}}=\coeff{1}{2}{|\Delta\Gamma||\epsilon|\over\cos\phi},\quad
\epsilon=|\epsilon|e^{-i\phi} \quad : \quad
\tan\phi={\Delta m\over {1\over2}|\Delta\Gamma|},
\end{equation}
with $|\epsilon|\approx2.2\times10^{-3}$ and $\phi\approx45^\circ$ the
``superweak" phase.

These equations are to be compared with the corresponding quantum-mechanical
equations of Ref.~\cite{peccei,emncpt} which are reviewed in
Appendix~\ref{app:A}. The parameters ${\delta}M$ and $\beta$ play similar
roles, although they appear with different relative signs in different places,
because
of the symmetry of $\delta\H$ as opposed to the antisymmetry of the
quantum-mechanical evolution matrix $H$. These differences are important for
the asymptotic limits of the density matrix, and its impurity. In our approach,
one can readily show that, at large $t$, $\rho$ decays exponentially to
\cite{emncpt}
\begin{equation}
\rho _L
\approx \left( \begin{array}{cc} 1 &
(|\epsilon | + i 2 {\widehat \beta }\cos\phi )e^{i\phi}  \\
(|\epsilon | - i 2 {\widehat \beta }\cos\phi )e^{-i\phi}  &
|\epsilon |^2 + {\widehat \gamma} -
4{\widehat \beta}^2 \cos ^2\phi - 4 {\widehat \beta} |\epsilon |\sin \phi
\end{array} \right)\ ,
\label{rhoL}
\end{equation}
where we have defined the following scaled variables
\begin{equation}
\widehat
\alpha=\alpha/|\Delta\Gamma|,\quad
\widehat\beta=\beta/|\Delta\Gamma|,
\quad \widehat\gamma=\gamma/|\Delta\Gamma|.
\label{abc}
\end{equation}
Conversely, if we look in the short-time limit
for a solution of the equations (\ref{rho11}) to (\ref{rho22}) with
${\rho}_{11}\ll  {\rho}_{12} \ll  {\rho}_{22}$, we find \cite{emncpt}
\begin{equation}
 \rho _S
\approx \left( \begin{array}{cc}
|\epsilon |^2 + {\widehat \gamma} -
4{\widehat \beta}^2 \cos ^2\phi + 4 {\widehat \beta} |\epsilon |\sin \phi
&
(|\epsilon | + i 2 {\widehat \beta }\cos\phi )e^{-i\phi}  \\
(|\epsilon | - i 2 {\widehat \beta }\cos\phi )e^{i\phi} &
1 \end{array} \right)\ .
\label{rhoS}
\end{equation}
These results are to be contrasted with those obtained within conventional
quantum mechanics
\begin{equation}
\rho_L\approx\left(
\begin{array}{cc}1&\epsilon^*\\ \epsilon&|\epsilon|^2\end{array}\right)\ ,
\qquad
\rho_S\approx\left(
\begin{array}{cc}|\epsilon|^2&\epsilon\\ \epsilon^*&1\end{array}\right)\ ,
\end{equation}
which, as can be seen from their vanishing determinant,\footnote{A pure
state will remain pure as long as $\Tr\rho^2=(\Tr\rho)^2$ \cite{EHNS}. In the
case of $2\times2$ matrices $\Tr\rho^2=(\Tr\rho)^2-2\det\rho$, and therefore
the purity condition is equivalently expressed as $\det\rho=0$.} correspond to
pure $K_L$ and $K_S$ states respectively. In contrast, $\rho_L,\rho_S$ in
Eqns.~(\ref{rhoL},\ref{rhoS}) describe mixed states. As mentioned in the
Introduction, the maximum possible order of magnitude for
$|\alpha|, |\beta|$ or $|\gamma |$ that we could expect theoretically is ${\cal
O}( E^2/M_{Pl})\sim {\cal O}((\Lambda _{\rm QCD}~{\rm or}~m_s)^2/M_{Pl})\sim
10^{-19}\GeV$ in the neutral kaon system.

To make a consistent phenomenological study of the various quantities discussed
above, it is essential to solve the coupled system of equations (\ref{rho11})
to (\ref{rho22}) for intermediate times. This requires approximations in order
to get analytic results \cite{Lopez}, as we discuss in the next section.

\section{Perturbation Theory}
\label{sec:perturbation}
The coupled set of differential equations (\ref{rho11}) to (\ref{rho22})
can be solved numerically to any desired degree of accuracy. However, it
is instructive and adequate for our purposes to solve these equations in
perturbation theory in $\widehat\alpha,\widehat\beta,\widehat\gamma$ and
$|\epsilon|$, so as to obtain convenient analytical approximations
\cite{Lopez}. Writing
\begin{equation}
 \rho_{ij}(t)=\rho^{(0)}_{ij}(t)+\rho^{(1)}_{ij}(t)+
\rho^{(2)}_{ij}(t)+\cdots
\label{perturb}
\end{equation}
where $\rho^{(n)}_{ij}(t)$ is proportional to
$\widehat\alpha^{p_\alpha}\widehat\beta^{p_\beta}
\widehat\gamma^{p_\gamma}|\epsilon|^{p_\epsilon}$, with
$p_\alpha+p_\beta+p_\gamma+p_\epsilon=n$, we obtain a set of differential
equations at each order in perturbation theory. To zeroth order we get
\begin{eqnarray}
\rho^{(0)}_{11}(t)&=&\rho_{11}(0)\,e^{-\Gamma_L t}\ ,\label{rho11^0}\\
\rho^{(0)}_{22}(t)&=&\rho_{22}(0)\,e^{-\Gamma_S t}\ ,\label{rho22^0}\\
\rho^{(0)}_{12}(t)&=&\rho_{12}(0)\,e^{-(\Gamma+i\Delta m t)}\ ,\label{rho12^0}
\label{zeroth}
\end{eqnarray}
where, in the interest of generality, we have left the initial conditions
unspecified. At higher orders the differential equations are of the form
\begin{equation}
\dot\rho^{(n)}_{ij}(t)=-A\rho^{(n)}_{ij}(t)
+\sum_{i'j'}\,\!\!'\rho^{(n-1)}_{i'j'}(t)
\end{equation}
where $\sum'$ excludes the $\rho_{ij}$ term. Multiplying by the integrating
factor $e^{At}$ one obtains
\begin{equation}
{d\over dt}
\left[e^{At}\rho^{(n)}_{ij}(t)\right]=
e^{At}\sum_{i'j'}\,\!\!'\rho^{(n-1)}_{i'j'}(t)
\end{equation}
which can be integrated in terms of the known
functions at the $(n-1)$-th order, and the initial condition
$\rho^{(n)}_{ij}(0)=0$, for $n\ge1$, \ie,
\begin{equation}
\rho^{(n)}_{ij}(t)=e^{-At}\int_0^t dt'\,
e^{At'}\,\sum_{i'j'}\,\!\!'\rho^{(n-1)}_{i'j'}(t')\ .
\label{solution}
\end{equation}

Following this straightforward (but tedious) procedure we obtain the
following set of first-order expressions
\begin{eqnarray}
\rho^{(1)}_{11}(t)&=&\rho_{22}(0)\widehat\gamma
\left[e^{-\Gamma_L t}-e^{-\Gamma_S t}\right]+{2|\epsilon|\over\cos\delta\phi}
|\rho_{12}(0)|\Bigl[e^{-\Gamma t}\cos(\Delta m t+\phi-\delta\phi-\phi_{12})
\nonumber\\
&&\qquad\qquad\qquad\qquad\qquad\qquad\qquad\qquad\qquad
-e^{-\Gamma_L t}\cos(\phi-\delta\phi-\phi_{12})\Bigr]\label{rho11^1}\\
\rho^{(1)}_{22}(t)&=&\rho_{11}(0)\widehat\gamma
\left[e^{-\Gamma_L t}-e^{-\Gamma_S t}\right]+{2|\epsilon|\over\cos\delta\phi}
|\rho_{12}(0)|\Bigl[e^{-\Gamma t}\cos(\Delta m t-\phi-\delta\phi-\phi_{12})
\nonumber\\
&&\qquad\qquad\qquad\qquad\qquad\qquad\qquad\qquad\qquad
-e^{-\Gamma_S t}\cos(\phi+\delta\phi+\phi_{12})\Bigr]\label{rho22^1}\\
\rho^{(1)}_{12}(t)&=&{2\widehat\alpha\over\tan\phi}|\rho_{12}(0)|e^{-\Gamma t}
\left[e^{-i\phi_{12}}\sin(\Delta mt)-(\Delta mt)
e^{-i\Delta mt+i\phi_{12}}\right]\nonumber\\
&&\qquad\qquad+{|\epsilon|\over\cos\delta\phi}\Biggl\{
\rho_{11}(0) e^{i(\delta\phi-\phi)}
\left[e^{-\Gamma_L t}-e^{-(\Gamma+i\Delta m)t}\right]\nonumber\\
&&\qquad\qquad\qquad\qquad\qquad+\rho_{22}(0) e^{i(\delta\phi+\phi)}
\left[e^{-\Gamma_S t}-e^{-(\Gamma+i\Delta m)t}\right]
\Biggr\}\label{rho12^1}
\end{eqnarray}
In these expressions $\phi_{12}={\rm Arg}\,[\rho_{12}(0)]$, and we have defined
\begin{equation}
\tan\delta\phi =-{2\widehat\beta\cos\phi\over|\epsilon|}\ .
\label{deltaphi}
\end{equation}
Note that generically {\em all} three parameters
($\widehat\alpha,\widehat\beta,\widehat\gamma$) appear to first order. However,
in the specific observables to be discussed below this is not necessarily the
case because of the particular initial conditions that may be involved. Thus,
these general expressions may be useful in the design of experiments that seek
to maximize the sensitivity to the CPT-violating parameters. To obtain the
expressions for $\Delta\rho$ and $\Sigma\rho$, one simply needs to insert the
appropriate set of initial conditions (Eq.~(\ref{InitialConditions})). Through
first order we obtain the following ready-to-use expressions:
\begin{eqnarray}
\Delta\rho^{(0+1)}_{11}(t)&=&{2|\epsilon|\over\cos\delta\phi}
\Bigl[-e^{-\Gamma t}\cos(\Delta m t+\phi-\delta\phi)
+e^{-\Gamma_L t}\cos(\phi-\delta\phi)\Bigr]\label{Deltarho11^0+1}\\
\Delta\rho^{(0+1)}_{22}(t)&=&{2|\epsilon|\over\cos\delta\phi}
\Bigl[-e^{-\Gamma t}\cos(\Delta m t-\phi-\delta\phi)
+e^{-\Gamma_S t}\cos(\phi+\delta\phi)\Bigr]\label{Deltarho22^0+1}\\
\Delta\rho^{(0+1)}_{12}(t)&=&-e^{-(\Gamma+i\Delta m)t}
-{2\widehat\alpha\over\tan\phi}e^{-\Gamma t}
\left[\sin(\Delta mt)-(\Delta mt)e^{-i\Delta mt}\right]\label{Deltarho12^0+1}\\
\Sigma\rho^{(0+1)}_{11}(t)&=&e^{-\Gamma_L t}+\widehat\gamma
\left[e^{-\Gamma_L t}-e^{-\Gamma_S t}\right]\label{Sigmarho11^0+1}\\
\Sigma\rho^{(0+1)}_{22}(t)&=&e^{-\Gamma_S t}+\widehat\gamma
\left[e^{-\Gamma_L t}-e^{-\Gamma_S t}\right]\label{Sigmarho22^0+1}\\
\Sigma\rho^{(0+1)}_{12}(t)&=&
{|\epsilon|\over\cos\delta\phi}\Biggl\{e^{i(\delta\phi-\phi)}
\left[e^{-\Gamma_L t}-e^{-(\Gamma+i\Delta m)t}\right]
+e^{i(\delta\phi+\phi)}\left[e^{-\Gamma_S t}-e^{-(\Gamma+i\Delta m)t}\right]
\Biggr\}\nonumber\\
&&\label{Sigmarho12^0+1}
\end{eqnarray}

For most purposes, first-order approximations suffice. However, in the
case of the $A_{2\pi}$ and $R_{2\pi}$ observables some second-order terms
in the expression for $\rho_{22}$ are required. For example,
$\Delta\rho_{22}^{(2)}$ introduces the first $\alpha$ dependence in the
numerator of $A_{2\pi}$, whereas $\Sigma\rho_{22}^{(2)}$ cuts off
the otherwise exponential growth with time of the numerator. The complete
second-order expressions for $\rho_{11,22,12}$ are collected in
Appendix~\ref{app:B}.

\section{Analytical Results}
\label{sec:analytical}
We now proceed to give explicit expressions for the temporal evolution of the
asymmetries $A_{2\pi},A_{3\pi}, A_{\rm T}, A_{\rm CPT}$, and $A_{\Delta m}$
that are possible objects of experimental study, in particular
by the CPLEAR collaboration
\cite{Guyot}.
\subsection{$A_{2\pi}$}
\label{sec:A2pi}
Following the discussion in section 2, one obtains for this asymmetry
\begin{equation}
A_{2\pi}(t)={\Delta\rho_{22}(t)\over\Sigma\rho_{22}(t)}\ ,
\label{Apipi}
\end{equation}
with $\Delta\rho_{22}$ and $\Sigma\rho_{22}$ given through first order in
Eqs.~(\ref{Deltarho22^0+1},\ref{Sigmarho22^0+1}); second-order contributions
can be obtained from Eq.~(\ref{rho22^2}). The result for the asymmetry, to
second order in the small parameters, can be written most concisely as
\begin{eqnarray}
A_{2\pi}(t)&=&\Biggl\{
\left[2|\epsilon|{\cos(\phi+\delta\phi)\over\cos\delta\phi}+\Delta X_1\right]
+e^{(\Gamma_S-\Gamma_L)t}\,\Delta X_2\nonumber\\
&&-e^{{1\over2}(\Gamma_S-\Gamma_L)t}\,
\left[{2|\epsilon|\over\cos\delta\phi}\,\cos(\Delta m t-\phi-\delta\phi)
+\Delta X_3\right]
\Biggr\}\nonumber\\
&&/\left\{[1-\widehat\gamma+\Sigma X_1]
+e^{(\Gamma_S-\Gamma_L)t}\,\left[\widehat\gamma+\Sigma X_2\right]
-e^{{1\over2}(\Gamma_S-\Gamma_L)t}\,\Sigma X_3 \right\}\ ,
\label{CPTx}
\end{eqnarray}
where the second-order coefficients $\Delta X_{1,2,3}$ and $\Sigma X_{1,2,3}$
are given by
\begin{eqnarray}
\Delta
X_1&=&2|\epsilon|\widehat\gamma\,{\cos(\phi+\delta\phi)\over\cos\delta\phi}
-8\widehat\alpha|\epsilon|\sin\phi\cos\phi{\sin(\phi+\delta\phi)
\over\cos\delta\phi}\\
\Delta X_2&=&
2|\epsilon|\widehat\gamma\,{\cos(\phi-\delta\phi)\over\cos\delta\phi}\\
\Delta X_3&=&
4|\epsilon|\widehat\gamma{\cos\phi\over\cos\delta\phi}\,
\cos(\Delta m t-\delta\phi)
+{4|\epsilon|\widehat\alpha\over\tan\phi}\,\sin(\Delta mt-\phi)\nonumber\\
&&-4|\epsilon|\widehat\alpha\,{\cos\phi\over\cos\delta\phi}\,
\left[{t|\Delta\Gamma|\over2\cos\phi}\,\cos(\Delta mt-\phi-\delta\phi)
-\cos(\Delta mt-2\phi-\delta\phi)\right] \label{DeltaX3}\\
\Sigma X_1&=&-\widehat\gamma^2+{2|\epsilon|^2\over\cos^2\delta\phi}\,
\left[\cos(2\delta\phi)+\cos(2\phi+2\delta\phi)
-{\cos(\phi-2\delta\phi)\over2\cos\phi}\right]\nonumber\\
&&+t|\Delta\Gamma|\left[-\widehat\gamma^2+|\epsilon|^2
{\cos(\phi+2\delta\phi)\over\cos\phi\cos^2\delta\phi}\right]\\
\Sigma X_2&=&\widehat\gamma^2+
|\epsilon|^2{\cos(\phi-2\delta\phi)\over\cos\phi\cos^2\delta\phi}\\
\Sigma X_3&=&{2|\epsilon|^2\over\cos^2\delta\phi}\left[
\cos(\Delta m t-2\delta\phi)+\cos(\Delta mt-2\phi-2\delta\phi)\right]
\end{eqnarray}
This form is useful when $\widehat\beta\ll|\epsilon|$, since then
$\delta\phi\approx0$. In the usual case (\ie,
$\widehat\alpha=\widehat\beta=\widehat\gamma=0$) we
obtain
\begin{equation}
A_{2\pi}(t)= {2|\epsilon|\cos\phi
-2|\epsilon|\,e^{{1\over2}(\Gamma_S-\Gamma_L)t}\,\cos(\Delta m t-\phi)\over
[1+\Sigma X^u_1]+e^{(\Gamma_S-\Gamma_L)t}\,\Sigma X^u_2
-e^{{1\over2}(\Gamma_S-\Gamma_L)t}\,\Sigma X^u_3}\ ,
\label{A2pi-usual}
\end{equation}
with
\begin{eqnarray}
\Sigma X^u_1&=&|\epsilon|^2\,[1+2\cos(2\phi)+t|\Delta\Gamma|]\\
\Sigma X^u_2&=&|\epsilon|^2\\
\Sigma X^u_3&=&4|\epsilon|^2\cos\phi\cos(\Delta mt-\phi)
\end{eqnarray}
Comparing the two cases we note the following:
\begin{enumerate}
\item The second line in Eq.~(\ref{CPTx}) shows that (to first order)
$\delta\phi\not=0$ changes the size of the interference pattern and shifts it.
\item The denominator in Eq.~(\ref{CPTx}) shows that necessarily
$\widehat\gamma\lsim\Sigma X_2\sim|\epsilon|^2$, or else the interference
pattern would be damped too soon. In fact, because of this upper limit one can
in practice neglect all terms proportional to $\widehat\gamma$ that appear
formally at second order, since they are in practice third order.
\item The effect of $\widehat\alpha$ is felt only at second order, through
$\Delta X_1$ and $\Delta X_3$, although it is of some relevance only in the
interference pattern ($\Delta X_3$).
\end{enumerate}
Some of the terms in Eq.~(\ref{CPTx}) can be written in a less concise
way which shows the effect of $\widehat\beta$ more explicitly
instead of it being buried inside $\delta\phi$. To first order, although
keeping the important second-order terms in $\Sigma X_2$, we can write
\begin{eqnarray}
A_{2\pi}(t)&\approx&\Biggl\{
2|\epsilon|\cos\phi+4\widehat\beta\sin\phi\cos\phi
-2\sqrt{|\epsilon|^2+4\widehat\beta^2\cos^2\phi}\,
e^{{1\over2}(\Gamma_S-\Gamma_L)t}\,
\cos(\Delta m t-\phi-\delta\phi)\Biggr\}\nonumber\\
&&/\left\{1+e^{(\Gamma_S-\Gamma_L)t}\,\left[\widehat\gamma+|\epsilon|^2
-4\widehat\beta^2\cos^2\phi-4\widehat\beta|\epsilon|\sin\phi\right]\right\}\ .
\label{CPTx2}
\end{eqnarray}
In this form one can readily see whether CP violation can in fact vanish, with
its effects mimicked by non-quantum-mechanical CPT violation. Setting
$|\epsilon|=0$ one needs to reproduce the interference pattern and also the
denominator. To reproduce the overall coefficient of the interference pattern
requires $2\widehat\beta\cos\phi\to\pm|\epsilon|$. The denominator then becomes
$\widehat\gamma-4\widehat\beta^2\cos^2\phi\to\widehat\gamma-|\epsilon|^2$ and
we also require $\widehat\gamma\to2|\epsilon|^2$. The fatal problem is that
$\delta\phi\to-{\rm sgn}(\widehat\beta){\pi\over2}$ and
the interference pattern is shifted significantly. This means that the effects
seen in the neutral kaon system, and conventionally interpreted as  CP
violation, indeed {\em cannot be due} to the CPT violation \cite{Lopez,HP}.

\begin{figure}[p]
\vspace{6in}
\includegraphics{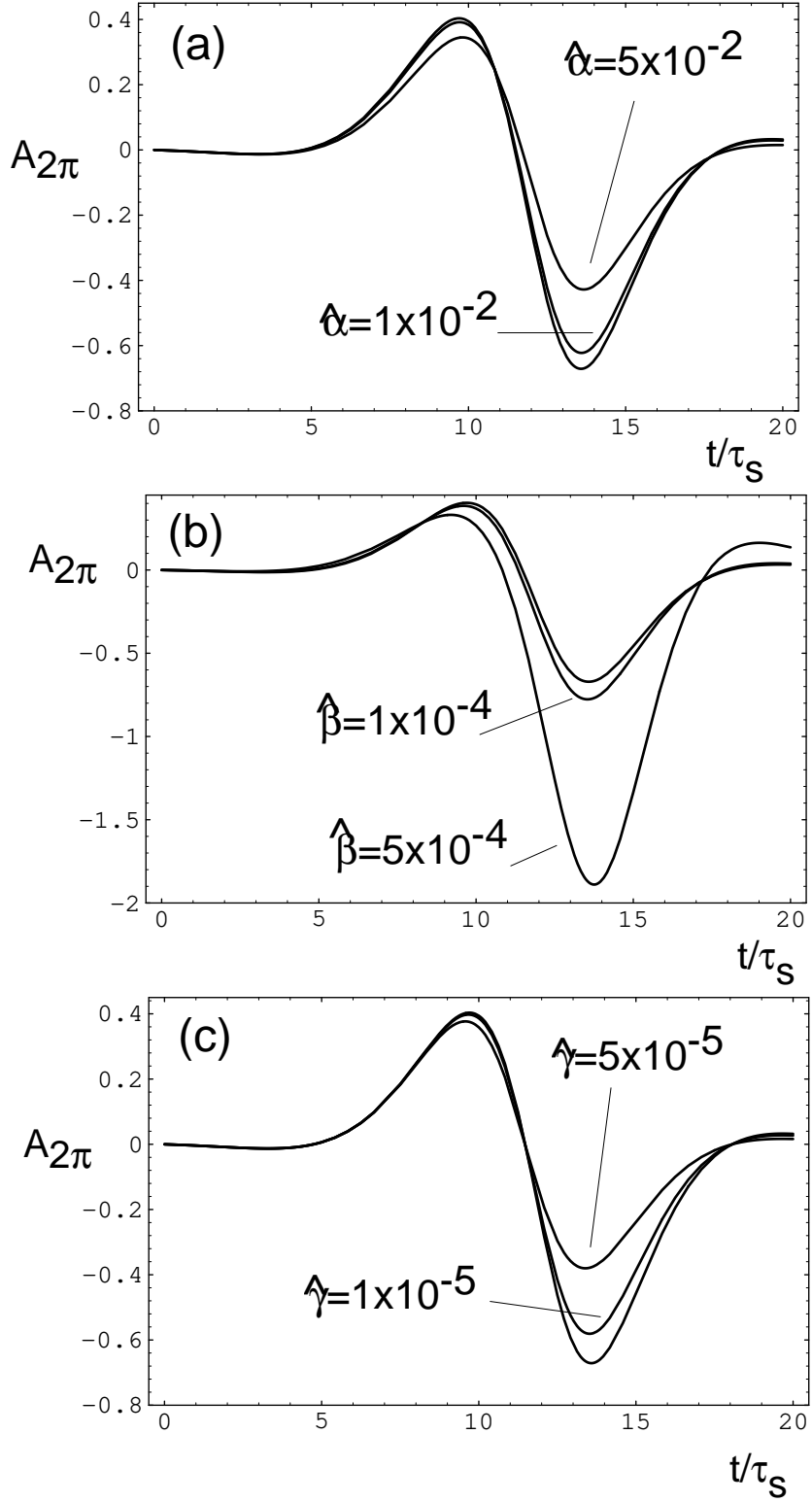}
\vspace{4.5cm}
\caption{The time-dependent asymmetry $A_{2\pi}$ for various choices of
the CPT-violating parameters: (a) dependence on $\widehat\alpha$, (b)
dependence on $\widehat\beta$, (c) dependence on $\widehat\gamma$. The
unspecified parameters are set to zero. The curve with no labels corresponds to
the standard case ($\widehat\alpha=\widehat\beta=\widehat\gamma=0$).}
\label{A2pi}
\end{figure}

Figure~\ref{A2pi} shows the effects on $A_{2\pi}(t)$  of varying
(a) ${\widehat \alpha }$, (b) ${\widehat \beta }$, and
(c) ${\widehat \gamma }$. We see that the intermediate-time
region $5 \lsim t/\tau _s \lsim 20 $ is particularly sensitive
to non-zero values of these parameters. The sensitivity
to ${\widehat \alpha }$ in Fig.~\ref{A2pi}(a) is considerably
less than that to ${\widehat \beta }$ in Fig.~\ref{A2pi}(b) and
${\widehat \gamma }$ in Fig.~\ref{A2pi}(c), which is reflected in the
magnitudes  of the indicative numerical bounds reported in
section~\ref{sec:Bounds}.

\subsection{$A_{3\pi}$}
\begin{figure}[p]
\vspace{6in}
\includegraphics{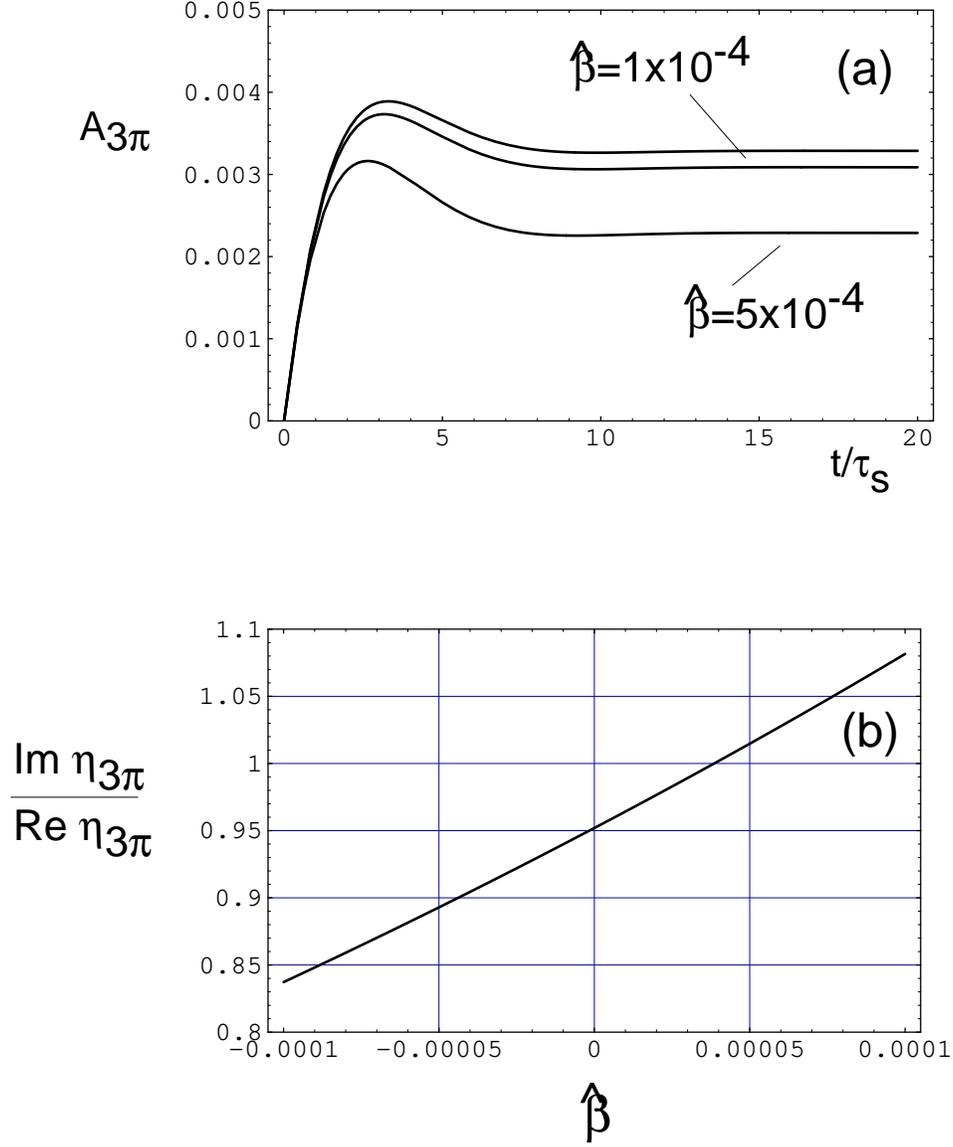}
\vspace{1.5cm}
\caption{The time-dependent asymmetry $A_{3\pi}$ for representative choices of
$\widehat\beta$ ($A_{3\pi}$ depends very weakly on
$\widehat\alpha,\widehat\gamma$). The top curve corresponds to the standard
case. Also shown is the ratio ${\rm Im}\eta_{3\pi}/{\rm
Re}\eta_{3\pi}=\tan(\phi-\delta\phi)$ as a function of $\widehat\beta$.}
\label{A3pi}
\end{figure}

Analogously, the formula for the $3\pi$ asymmetry is
\begin{equation}
A_{3\pi}(t) =
\frac{{\rm Tr}[O_{3\pi}\, {\bar \rho}(t)]-{\rm Tr}[O_{3\pi}\, \rho(t)]}
{\rm{Tr}[O_{3\pi}\, {\bar \rho}(t)]+ {\rm Tr}[O_{3\pi}\, \rho(t)]} \qquad ;
\qquad O_{3\pi} \propto\left(\begin{array}{cc}
1 & 0 \\
0 & 0 \end{array}\right)\ ,
\label{asim3p}
\end{equation}
from which one immediately obtains
\begin{equation}
A_{3\pi}(t) =
\frac{\Delta \rho _{11}(t) }{\Sigma \rho _{11} (t) }\ .
\label{3p1}
\end{equation}
To first order in the small parameters, $\Delta\rho_{11}$ and $\Sigma\rho_{11}$
are given in Eqns.~(\ref{Deltarho11^0+1},\ref{Sigmarho11^0+1}). This asymmetry
can therefore be expressed~as
\begin{eqnarray}
A_{3\pi}(t)&=&{2|\epsilon|{\cos(\phi-\delta\phi)\over\cos\delta\phi}-
{2|\epsilon|\over\cos\delta\phi}e^{-\coeff{1}{2}(\Gamma_S-\Gamma_L)t}
\cos(\Delta mt+\phi-\delta\phi)\over
1+\widehat\gamma-\widehat\gamma e^{-(\Gamma_S-\Gamma_L)t}}\\
&\approx&\left[2|\epsilon|\cos\phi-4\widehat\beta\sin\phi\cos\phi\right]
-2e^{-{1\over2}(\Gamma_S-\Gamma_L)t}
\left[{\rm Re}\eta_{3\pi}\cos\Delta mt - {\rm Im}\eta_{3\pi}\sin\Delta
 mt\right]\ ,\nonumber
\label{3p2}
\end{eqnarray}
where, to facilitate contact with experiment, in the second form we have
neglected the $\widehat\gamma$ contribution, expressed $\delta\phi$ in terms of
$\widehat\beta$ (\ref{deltaphi}), and defined
\begin{equation}
{\rm
Re}\eta_{3\pi}=|\epsilon|\cos\phi-2\widehat\beta\sin\phi\cos\phi,\quad
{\rm Im}\eta_{3\pi}=|\epsilon|\sin\phi+2\widehat\beta\cos^2\phi\ .
\label{3p3}
\end{equation}
In the CPLEAR experiment, the time-dependent decay asymmetry into
$\pi^0\pi^+\pi^-$ is measured \cite{Guyot}, and the data is fit to obtain
the best values for ${\rm Re}\eta_{3\pi}$ and ${\rm Im}\eta_{3\pi}$. It would
appear particularly useful to determine the ratio of these two parameters, so
that a good fraction of the experimental uncertainties drops out. In the
standard CP-violating scenario, the ratio is ${\rm Im}\eta_{3\pi}/{\rm
 Re}\eta_{3\pi}=\tan\phi\approx1$, whereas in our scenario it is
\begin{equation}
{{\rm Im}\eta_{3\pi}\over{\rm Re}\eta_{3\pi}}=
{|\epsilon|\sin\phi+2\widehat\beta\cos^2\phi
\over |\epsilon|\cos\phi-2\widehat\beta\sin\phi\cos\phi}=\tan(\phi-\delta\phi)\
{}.
\label{3p4}
\end{equation}
It is apparent from the above formulae that $A_{3\pi}$ is much more sensitive
to ${\widehat \beta }$ that to ${\widehat \alpha }$ or ${\widehat \gamma}$.
This sensitivity of $A_{3\pi}$ to ${\widehat \beta}$ is shown
in Fig.~\ref{A3pi}(a), and that of $({\rm Im}~\eta_{3\pi} /{\rm
Re}~\eta_{3\pi})$ in Fig.~\ref{A3pi}(b).

As already mentioned in Sec.~\ref{sec:observables}, additional information
may be obtained from $\pi^+\pi^-\pi^0$ decays by observing the difference
between the rates for $\pi^+\pi^-\pi^0$ decays with
$m(\pi^+\pi^0)>m(\pi^-\pi^0)$ and $m(\pi^+\pi^0)<m(\pi^-\pi^0)$~\cite{nakada},
represented by $O_{3\pi}^{\rm int}$ (\ref{XI},\ref{Y}). This division of the
final-state phase space into two halves is not CP-invariant, and hence enables
one to measure interference between the CP-even $I_{3\pi}=2$ and  CP-odd
$I_{3\pi}=1$ final states. Defining
\begin{equation}
A_{3\pi}^{\rm int} =
\frac{{\rm Tr} [O_{3\pi}^{\rm int}{\overline\rho(t)}]
-{\rm Tr}[O_{3\pi}^{\rm int}\rho( t)]}
{{\rm Tr}[O_{3\pi}^{\rm int}{\overline\rho(t)}]+{\rm Tr}[O_{3\pi}^{\rm int}
\rho(t)]}\ ,
\label{Z}
\end{equation}
we obtain the formula
\begin{equation}
A_{3\pi}^{\rm int}=\frac{{\rm Re}\Delta\rho _{12}}{{\rm Re}\Sigma\rho_{12}}\ .
\label{ZZ}
\end{equation}
To first order in small parameters, we find
\begin{equation}
A_{3\pi}^{\rm int}={-e^{-\Gamma t}\left[\cos\Delta mt+
{2\widehat\alpha\over\tan\phi}(\sin\Delta mt-(\Delta m t)\cos\Delta mt)\right]
\over
{|\epsilon|\over\cos\delta\phi}
\left[\cos(\phi-\delta\phi)e^{-\Gamma_L t}+\cos(\phi+\delta\phi)e^{-\Gamma_S t}
-2\cos\phi e^{-\Gamma t}\cos(\Delta mt-\delta\phi)\right]}
\label{ZZZ}
\end{equation}
Note that $A_{3\pi}^{\rm int}\to\infty$ for $t\to0$. In the CPT-conserving
case this observable becomes
\begin{equation}
A_{3\pi}^{\rm int}\to{-e^{-\Gamma t}\cos\Delta mt
\over|\epsilon|\cos\phi
\left[e^{-\Gamma_L t}+e^{-\Gamma_S t}-2e^{-\Gamma t}\cos\Delta mt\right]}
\label{ZZZ2}
\end{equation}
We see that this observable is sensitive to $\widehat\alpha$ (see the numerator
of (\ref{ZZZ})), and to $\widehat\beta$ via $\delta\phi$. The sensitivity to
$\widehat\alpha$ may supplement usefully the information obtainable from the
$A_{\Delta m}$ measurement discussed in section~\ref{sec:ADeltam}.

\subsection{$A_{\rm T}$}
\begin{figure}[p]
\vspace{6in}
\includegraphics{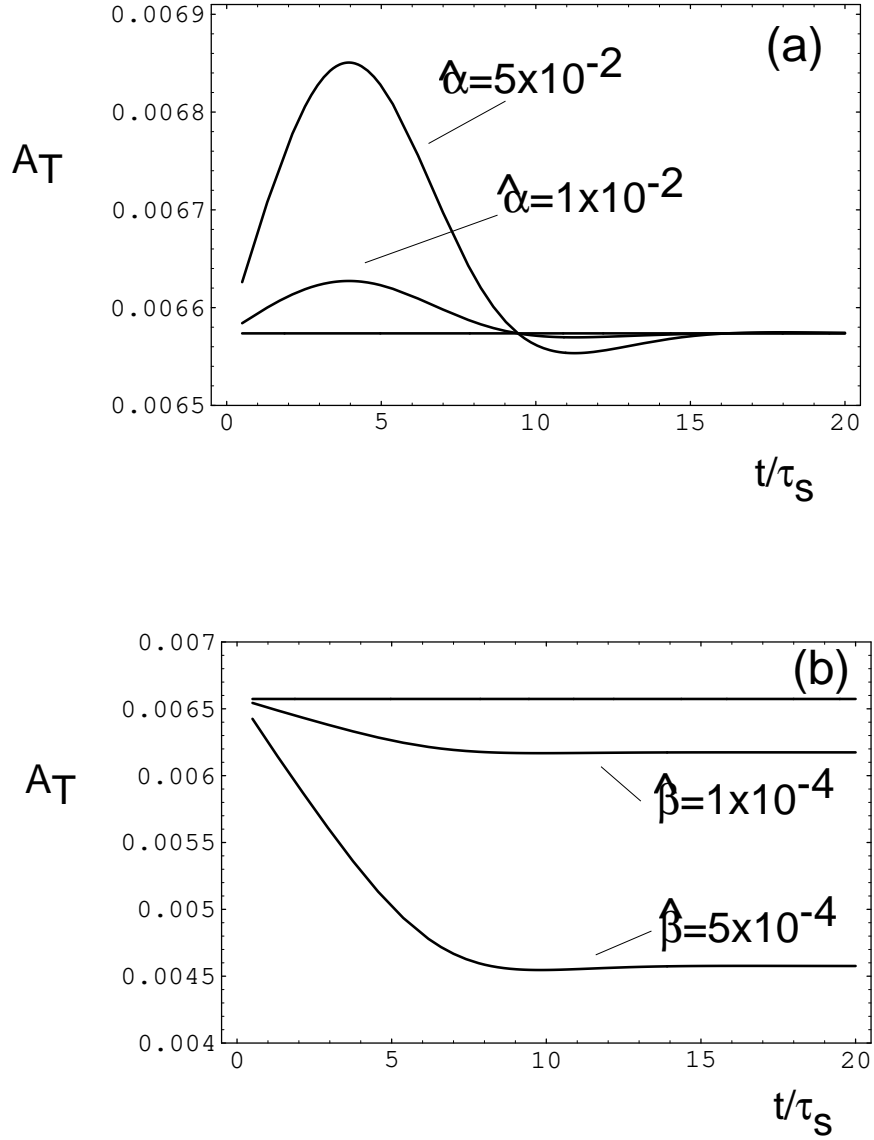}
\vspace{1.5cm}
\caption{The time-dependent asymmetry $A_{\rm T}$ for representative choices of
(a) $\widehat\alpha$ ($\widehat\beta=0$) and (b) $\widehat\beta$
($\widehat\alpha=0$). The dependence on $\widehat\gamma$ is negligible. The
flat line corresponds to the standard case.}
\label{AT}
\end{figure}

The formula for this asymmetry, as obtained by applying the formalism of
section~\ref{sec:observables}, assumes the form
\begin{equation}
A_{\rm T}={\Delta\rho_{11}+\Delta\rho_{22}+2{\rm Re}\,\Sigma\rho_{12}
\over \Sigma\rho_{11}+\Sigma\rho_{22}+2{\rm Re}\,\Delta\rho_{12}}\ ,
\label{Atf}
\end{equation}
with the first-order expressions for $\Delta\rho_{11,22,12}$ and
$\Sigma\rho_{11,22,12}$ given in
Eqns.~(\ref{Deltarho11^0+1})--(\ref{Sigmarho12^0+1}).
In the usual non-CPT-violating case one finds, to first order, the following
exactly time-independent result
\begin{equation}
A_{\rm T}=4|\epsilon|\cos\phi\ ,
\label{CPTx=0}
\end{equation}
as expected \cite{Guyot}. In the CPT-violating case, to first order, one finds
a time-dependent expression
\begin{eqnarray}
A_{\rm T}&=&{4|\epsilon|\over\cos\delta\phi}\nonumber\\
&&\left\{{e^{-\Gamma_L t}\cos(\phi-\delta\phi)+e^{-\Gamma_S
t}\cos(\phi+\delta\phi)
-2e^{-\Gamma t}\cos(\Delta m t-\delta\phi)\cos\phi
\over
e^{-\Gamma_L t}(1+2\widehat\gamma)+e^{-\Gamma_S t}(1-2\widehat\gamma)
-2e^{-\Gamma t}
[\cos\Delta m t+{2\widehat\alpha\over\tan\phi}(\sin\Delta m t
-\Delta m t\cos\Delta m t)]
}\right\}\nonumber\\
\label{At}
\end{eqnarray}
which aymptotes to
\begin{equation}
A_{\rm T}\to
 {4|\epsilon|\cos(\phi-\delta\phi)\over\cos\delta\phi
 (1+2\widehat\gamma)}
={4|\epsilon|\cos\phi-8\widehat\beta\sin\phi\cos\phi\over1+2\widehat\gamma}\ .
\label{Ata}
\end{equation}
The sensitivity of $A_{\rm T}$ to ${\widehat \alpha}$ and ${\widehat \beta }$
are illustrated in Fig.~\ref{AT}(a) and Fig.~\ref{AT}(b), respectively.
We see that the sensitivity to ${\widehat \alpha }$  is again less than
that to ${\widehat \beta }$, and is restricted to $t/\tau _s \lsim 15 $,
whereas the greater sensitivity to ${\widehat \beta }$ persists to
large $t$, as in Eq.~(\ref{Ata}), where the corresponding (utterly negligible)
sensitivity to $\widehat\gamma$ can be inferred.

\subsection{$A_{\rm CPT}$}
\label{sec:A_CPT}
Following the discussion in section~\ref{sec:observables}, the formula for this
observable, as defined by the CPLEAR Collaboration \cite{Guyot}, is given by
Eq.~(\ref{asym}) with $f=\pi^-l^+\nu$ and $\bar f=\pi^+l^-\bar\nu$. We obtain
\begin{equation}
A_{\rm CPT}={\Delta\rho_{11}+\Delta\rho_{22}-2{\rm Re}\,\Sigma\rho_{12}
\over \Sigma\rho_{11}+\Sigma\rho_{22}-2{\rm Re}\,\Delta\rho_{12}}\ .
\label{Acptf}
\end{equation}
To first order, in both the CPT-conserving and CPT-violating cases, we find
\begin{equation}
A_{\rm CPT}=0\ .
\label{Acpt=0}
\end{equation}
To second order, the terms in the numerator of Eq.~(\ref{Acptf}) can be
written most succinctly in the long-time limit. With the help of the
expressions in Appendix~\ref{app:B} we obtain
\begin{eqnarray}
\Delta\rho_{11}^{(2)}&\to&-2|\epsilon|\widehat\gamma\cos\phi
+8|\epsilon|\widehat\alpha\cos\phi\sin^2\phi
+4\widehat\beta\widehat\gamma\sin\phi\cos\phi
+16\widehat\alpha\widehat\beta\sin\phi\cos^3\phi\nonumber\\
\Delta\rho_{22}^{(2)}&\to&2|\epsilon|\widehat\gamma\cos\phi
-4\widehat\beta\widehat\gamma\sin\phi\cos\phi\\
{\rm
Re}\,\Sigma\rho_{12}^{(2)}&\to&4|\epsilon|
\widehat\alpha\cos\phi\sin^2\phi
+8\widehat\alpha\widehat\beta\sin\phi\cos^3\phi\nonumber
\end{eqnarray}
which show that in the long-time limit $A_{\rm CPT}=0$ also to second order.
In fact, some algebra shows that $A_{\rm CPT}=0$ through second order for {\em
all} values of $t$. This result implies that $|A_{\rm CPT}|\lsim10^{-6}$ and
thus is unobservably small.

We point out that this result is a {\em quite distinctive signature} of the
modifications of the quantum mechanics proposed in Ref.~\cite{EHNS,
emncpt}, since in the case of quantum-mechanical violation of CPT symmetry
\cite{peccei} there is a non-trivial change in $A_{\rm CPT}$, proportional to
the CPT-violating parameters $\delta M$ and $\delta\Gamma$. Indeed, in
Appendix~\ref{app:A} we obtain the following first-order asymptotic result
\begin{equation}
A_{\rm CPT}^{\rm QM}\to4\sin\phi\cos\phi\,\widehat{\delta M}
+2\cos^2\phi\,\widehat{\delta\Gamma}\ ,
\label{acptqm}
\end{equation}
written in terms of the scaled variables. Part of the reason for this
difference is the different role played by $\delta M$ as compared to the
$\beta$ parameter in the formalism of Ref.~\cite{EHNS},
as discussed in detail
in Ref.~\cite{emncpt}. In particular, there are important sign
differences between the ways that $\delta M$ and $\beta$ appear in the two
formalisms, that cause the suppression to second order of any
quantum-mechanical-violating effects in $A_{\rm CPT}$, as opposed to the
conventional quantum mechanics case.

\subsection{$A_{\Delta m}$}
\label{sec:ADeltam}
Following Ref.~\cite{Guyot}, one can define $A_{\Delta m}$ as
\begin{equation}
A_{\Delta m}={R(K^0\to\pi^+)+R(\bar K^0\to\pi^-)-R(\bar K^0\to\pi^+)
-R(K^0\to\pi^-)\over R(K^0\to\pi^+)+R(\bar K^0\to\pi^-)+R(\bar
K^0\to\pi^+)
+R(K^0\to\pi^-)}
\label{Adeltmadef}
\end{equation}
in an obvious short-hand notation for the final states of the semileptonic
decays, where only the pion content is shown explicitly.  In the formalism
of section~\ref{sec:observables}, this expression becomes
\begin{equation}
A_{\Delta m}={2{\rm
Re}\Delta\rho_{12}\over\Sigma\rho_{11}+\Sigma\rho_{22}}\ .
\label{Adeltam}
\end{equation}
The first-order expression in the usual non-CPT violating case is
\begin{equation}
 A_{\Delta m}=-{2e^{-\Gamma t}\cos\Delta m t\over e^{-\Gamma_L
t}+e^{-\Gamma_S t}}\ ,
\label{Adeltam1}
\end{equation}
as obtained in Ref.~\cite{Guyot}. In the CPT-violating case to first order, as
Eqs.~(\ref{Deltarho12^0+1},\ref{Sigmarho11^0+1},\ref{Sigmarho22^0+1}) show,
neither $|\epsilon|$ nor $\widehat\beta$ come in,
and we obtain
\begin{equation}
A_{\Delta m}=-{2e^{-\Gamma t}
\left[\cos\Delta m t+{2\widehat\alpha\over\tan\phi}
(\sin\Delta mt-\Delta mt\cos\Delta m t)\right]\over
e^{-\Gamma_L t}(1+2\widehat\gamma)+e^{-\Gamma_S t}(1-2\widehat\gamma)}\ .
\label{Adeltam2}
\end{equation}
Since $\widehat\gamma$ is negligible, this observable provides an {\em
exclusive} test of $\widehat\alpha$.

\begin{figure}[p]
\vspace{6in}
\includegraphics{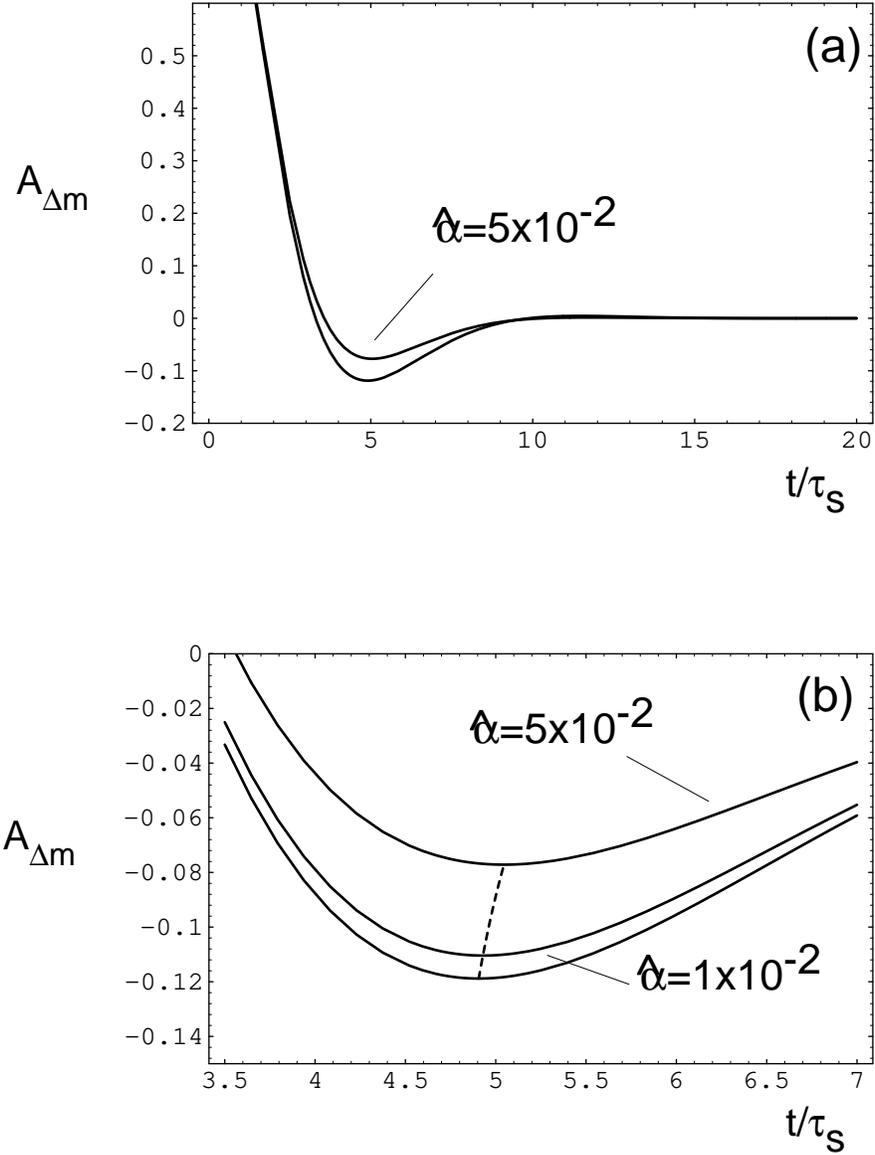}
\vspace{1.5cm}
\caption{The time-dependent asymmetry $A_{\Delta m}$ for representative choices
of $\widehat\alpha$ ($\widehat\beta=\widehat\gamma=0$). This asymmetry depends
most sensitively only on $\widehat\alpha$. In both panels, the bottom curve
corresponds to the standard case. In the detail (b), the dashed line indicates
the location of the minimum as $\widehat\alpha$ is varied.}
\label{ADeltam}
\end{figure}

In the case of no CPT violation, the observable $A_{\Delta m}$ has a minimum
for $\tan\Delta mt=-\Gamma/\Delta m\approx -\coeff{1}{2}|\Delta\Gamma|/\Delta
m=-1/\tan\phi$. Since $\tan\phi\approx1$, the minimum occurs for
$(t/\tau_s)_{\rm min}\approx3\pi/2$. In the CPT-violating case,
Eq.~(\ref{Adeltam2}) can be
rewritten as
\begin{equation}
 A_{\Delta m}=-{2e^{-\Gamma t}C_{\Delta m}\cos(\Delta m t-\phi_{\Delta m})\over
e^{-\Gamma_L t}(1+2\widehat\gamma)+e^{-\Gamma_S t}(1-2\widehat\gamma)}\ .
\label{Adeltam3}
\end{equation}
with
\begin{equation}
\tan\phi_{\Delta m}={2\widehat\alpha/\tan\phi\over1-\widehat\alpha
t|\Delta\Gamma|}\, ,\qquad C_{\Delta m}={1-\widehat\alpha t|\Delta\Gamma|
\over\cos\phi_{\Delta m}}\ .
\label{Adeltam4}
\end{equation}
Since the minimum occurs for $t|\Delta\Gamma|\sim5$, for small values of
$\widehat\alpha$ one can neglect the
time-dependent pieces in $\phi_{\Delta m}$ and $C_{\Delta m}$. The new minimum
condition for $A_{\Delta m}$ is then modified to $\tan(\Delta m t-\phi_{\Delta
m})\approx-1/\tan\phi$, and thus the minimum is shifted to
\begin{equation}
(t/\tau_s)_{\rm min}\approx \coeff{3\pi}{2}+4\widehat\alpha\ ,
\label{Adeltam5}
\end{equation}
for small values of $\widehat\alpha$. A similar test for $\widehat\alpha$ was
proposed in Ref.~\cite{Lopez}, where it was based on the traditional
semileptonic decay charge asymmetry parameter $\delta(t)$ \cite{emncpt}.
However, to first order that observable depends also on $|\epsilon|$ and
$\widehat\beta$, and as such it is {\em not} a direct test of $\widehat\alpha$,
as opposed to the one proposed here.
Figure~\ref{ADeltam} exhibits the sensitivity of $A_{\Delta m}$
to ${\widehat \alpha }$, including (a) the general dependence
in the interference region and (b) the detailed location of the minimum
as ${\widehat \alpha }$ is varied.

\section{Regeneration}
\label{sec:regeneration}
\subsection{Simplified Thin-Regenerator Case}
Regeneration involves the coherent scattering of a $K^0$ or ${\bar K}^0$
off a nuclear target, which we assume can be described using the normal
framework of quantum field theory and quantum mechanics. Thus we
describe it by an effective Hamiltonian which takes the form
\begin{equation}
  \Delta H = \left( \begin{array}{cc}T&0\\0&{\overline T}
     \end{array}\right)
\label{4.6.1}
\end{equation}
in the $(K^0, {\bar K}^0)$ basis, where
\begin{equation}
T=\frac{2\pi N}{m_K}\, {\cal M}\,,\qquad
\overline T=\frac{2\pi N}{m_K}\,\overline{\cal M}
\label{4.6.2}
\end{equation}
with ${\cal M} = \vev{K^0|A|K^0}$ the forward $K^0$-nucleus scattering
amplitude (and analogously for $\overline{\cal M}$), and $N$ is the nuclear
regenerator density. We can rewrite $\Delta H$ (\ref{4.6.1}) in the $K_{1,2}$
basis as
\begin{equation}
 \Delta H = \left( \begin{array}{cc}
 T + {\overline T}&  T - {\overline T}  \\
  T - {\overline T}& T + {\overline T}
    \end{array}\right)\ ,
\label{4.6.3}
\end{equation}
which can in principle be included as a contribution to $H$ in the density
matrix equation:
\begin{equation}
   \partial _t \rho = -i [ H, \rho] + i\delta\H\rho\ ,
\label{4.6.4}
\end{equation}
where $\delta\H$ represents the possible CPT- and QM-violating term.

It may be adequate as a first approximation to treat the regenerator as
very thin, in which case we may use the impulse approximation,
and the regenerator changes  $\rho$ by an amount
\begin{equation}
  \delta \rho = -i[\Delta {\cal H}, \rho ]\ ,
\label{4.6.5}
\end{equation}
where
\begin{equation}
  \Delta {\cal H} = \int dt \Delta H\ .
\label{4.6.6}
\end{equation}
Writing
\begin{equation}
\rho=\left(
\begin{array}{cc}\rho_{11}&\rho^*_{12}\\ \rho_{12}&\rho_{22}\end{array}
\right)\ ,
\end{equation}
in this approximation we obtain
\begin{equation}
\delta \rho = -i \Delta T
 \left( \begin{array}{cc}
2i{\rm Im}\rho_{12}&-\rho_{11}+\rho_{22}\\
\rho_{11}-\rho_{22}&-2i{\rm Im}\rho_{12}\end{array}\right)\ ,
\label{4.6.7}
\end{equation}
where
\begin{equation}
\Delta T \equiv \int dt (T - {\overline T} )\ .
\label{4.6.8}
\end{equation}
This change in $\rho$ enables the possible CPT- and QM-violating terms in
(\ref{4.6.4}) to be probed in a new way. Consider the idealization that the
neutral $K$ beam is already in a $K_L$ state (Eq.~(\ref{rhoL})):
\begin{equation}
  \rho = \rho _L \approx \left( \begin{array}{cc}
1&  \epsilon ^* + B^*  \\
  \epsilon + B &  |\epsilon |^2 + C
 \end{array}\right)
\label{4.6.9}
\end{equation}
where
\begin{equation}
B = -i2\widehat\beta\cos\phi\, e^{-i\phi}\qquad ; \qquad
C=\widehat\gamma-4\widehat\beta^2\cos^2\phi-4
\widehat\beta|\epsilon|\sin\phi
\label{4.6.10}
\end{equation}
Substituting Eqs.~(\ref{4.6.9},\ref{4.6.10}) into Eq.~(\ref{4.6.7}),
we find that in the  joint large-$t$ and impulse approximations
\begin{equation}
    \rho + \delta \rho =
  \left( \begin{array}{cc}
 1  + 2\Delta T {\rm Im}(\epsilon + B )
&  \epsilon ^* + B^*
  + i (1 - |\epsilon |^2 - C)\Delta T  \\
  \epsilon + B
  - i (1 - |\epsilon |^2 - C)\Delta T
&  |\epsilon |^2 + C  - 2\Delta T {\rm Im}(\epsilon + B)
 \end{array}\right)\ .
\label{4.6.11}
\end{equation}
We see that the usual semileptonic decay asymmetry observable
\begin{equation}
     O_{\pi^-l^+\nu} -
     O_{\pi^+l^-\bar\nu} = \left(\begin{array}{cc}
0 & 2 \\
2 & 0 \end{array}\right) \qquad ,
\label{4.6.12}
\end{equation}
which measures ${\rm Re}(\epsilon + B)$ in the case without the
regenerator, receives no contribution from the regenerator (\ie, $\Delta T$
cancels out in the sum of the off-diagonal elements). On the other hand, there
is a new contribution to the value of $R_{2\pi}=R(K_L\to2\pi)={\rm
Tr}[O_{2\pi}\rho]=\rho_{22}$, namely
\begin{equation}
R_{2\pi} = |\epsilon|^2
+\widehat\gamma-4\widehat\beta^2\cos^2\phi-4
\widehat\beta|\epsilon|\sin\phi
-2\Delta T {\rm Im}(\epsilon + B)\ .
\label{4.6.15}
\end{equation}
The quantity ${\rm Im}(\epsilon + B)$ was not accessible directly to the
observable $R_{2\pi}$ in the absence of a regenerator.
Theoretically, the phases of $\epsilon$ and $B$ (\ref{4.6.10})
are fixed, \ie,
\begin{equation}
{\rm Im}(\epsilon +
B)=-|\epsilon|{\sin(\phi-\delta\phi)\over\cos\delta\phi}=
-|\epsilon|\sin\phi-2\widehat\beta\cos^2\phi\ .
\label{4.6.16}
\end{equation}
Nevertheless, this phase prediction should be checked, so the regenerator makes
a useful addition to the physics programme.

The above analysis is oversimplified, since the impulse approximation may not
be sufficiently precise, and the neutral $K$ beam is not exactly in a
$K_L$ state. Moreover, the result in Eq.~(\ref{4.6.11}) is valid {\em only}
at the time the beam emerges from the regenerator. However, this simple example
may serve to illustrate the physics interest of measurements using a
regenerator. We next generalize the analysis to include a general neutral $K$
beam encountering a thin regenerator, with the full time dependence after
leaving the regenerator.

\subsection{Detailed Regenerator Tests}
To make contact with the overall discussion in this paper, we envision
the following scenario:
\begin{description}
\item (i) Pure $K^0,\bar K^0$ beams are produced at time $t=0$,
corresponding to initial density matrices $\rho_0$ and $\bar\rho_0$,
respectively.
\item (ii) These beams are described by density matrices $\rho(t)$ and
$\bar\rho(t)$, and evolve with time as described in
Section~\ref{sec:observables}, until a time $t=t_r$ where they are described
by $\rho(t_r)$ and $\bar\rho(t_r)$.
\item (iii) At $t=t_r$ a thin regenerator is encountered.\footnote{For
simplicity we assume that the regenerator is encountered at the same $\Delta
t=t_r$ after production for {\em all} beam particles. In specific experimental
setups our expressions would need to be folded with appropriate geometrical
functions.} In our thin-regenerator approximation (described in the previous
subsection), at $t=t_r$ suddenly the density matrices receive an additional
contribution $\delta\rho(t_r)$ and $\delta\bar\rho(t_r)$, according to
Eq.~(\ref{4.6.7}).
\item (iv) For $\tau=t-t_r\ge0$, the beams are described by density matrices
$\rho^r(\tau)$ and $\bar\rho^r(\tau)$, which again evolve as described in
Section~\ref{sec:observables}, but this time with initial conditions
$\rho^r(0)=\rho(t_r)+\delta\rho(t_r)$ and
$\bar\rho^r(0)=\bar\rho(t_r)+\delta\bar\rho(t_r)$.
\end{description}
In this context, we consider two kinds of tests. In a CPLEAR-like scenario,
the identity of the beam is known irrespective of the presence of the
regenerator, and thus a measurement of $A^r_{2\pi}(\tau)$, \ie, $A_{2\pi}$
after the thin regenerator is traversed, appears feasible. The second test
is reminiscent of the Fermilab experiments, where the experimental setup is
such that $t_r\gg\tau_S$, and the beam is in a $K_L$ state. After the
regenerator is traversed one then measures $R_{2\pi}$ in the interference
region.

Before embarking on elaborate calculations, we should perhaps quantify our
``thin-regenerator" criterion. For the impulse approximation to be valid,
$\delta\rho$ in Eq.~(\ref{4.6.7}) should not change $\rho$ by too much. Since
the entries in $\rho$ are typically ${\cal O}(10^{-3})$ or smaller, we should
demand that $\Delta T$ be a reasonably small number. Let us estimate $\Delta
T=\int dt(T-\bar T)$. Assuming ${\cal M}-\overline{\cal M}\sim1/m_\pi$ and
relativistic kaons
we obtain
\begin{equation}
\Delta T\approx \coeff{1}{30}\, {\rm thickness}\,[{\rm cm}]\ {\rm
density}\,[{\rm g/cm^3}]\ ,
\label{estimate}
\end{equation}
and thus a ``thin" regenerator should have a thickness $\lsim{\cal O}(1\,{\rm
cm})$. This estimate appears reasonable when considering that in the 2~ns or so
that the beams are usually observed (about $20\tau_S$), they travel
$\sim60\,{\rm cm}$. Such a regenerator could conceivably be installed in an
upgraded CPLEAR experiment. In the Fermilab E731~\cite{E731} and
E773~\cite{E773,E773-prl} experiments the regenerators used are much thicker,
and the validity of our approximation is unclear.
\subsubsection{$A^r_{2\pi}$}
We start with $A^r_{2\pi}=\Delta\rho^r_{22}/\Sigma\rho^r_{22}$, where \eg,
$\Delta\rho^r_{22}(\tau)$ is given by $\rho_{22}(\tau)$ in
Eqns.~(\ref{rho22^0},\ref{rho22^1}) with
$\rho(0)\to\Delta\rho^r(0)=\Delta\rho(t_r)+\delta(\Delta\rho)(t_r)$, and
$\delta(\Delta\rho)$ given in Eq.~(\ref{4.6.7}) with $\rho\to\Delta\rho$. We
obtain
\begin{eqnarray}
\Delta\rho^r_{22}(\tau)&=&
\Biggl\{[\Delta\rho_{22}(t_r)-2\Delta T{\rm Im}\Delta\rho_{12}(t_r)]-
\widehat\gamma\,[\Delta\rho_{11}(t_r)+2\Delta T{\rm Im}\Delta\rho_{12}(t_r)]
\nonumber\\
&&-2|\epsilon|\,{\cos(\phi+\delta\phi+\Delta\phi_{12})\over\cos\delta\phi}\,
|\Delta\rho^r_{12}(0)|\Biggr\}e^{-\Gamma_S \tau}\nonumber\\
&&+\widehat\gamma\,[\Delta\rho_{11}(t_r)+2\Delta T{\rm Im}\Delta\rho_{12}(t_r)]
e^{-\Gamma_L \tau}\nonumber\\
&&+{2|\epsilon|\over\cos\delta\phi}|\Delta\rho^r_{12}(0)|e^{-\Gamma\tau}
\cos(\Delta m\tau-\phi-\delta\phi-\Delta\phi_{12})\label{Deltarho22^r}\ ,\\
\Sigma\rho^r_{22}(\tau)&=&
\Biggl\{[\Sigma\rho_{22}(t_r)-2\Delta T{\rm Im}\Sigma\rho_{12}(t_r)]-
\widehat\gamma\,[\Sigma\rho_{11}(t_r)+2\Delta T{\rm Im}\Sigma\rho_{12}(t_r)]
\nonumber\\
&&-2|\epsilon|\,{\cos(\phi+\delta\phi+\Sigma\phi_{12})\over\cos\delta\phi}\,
|\Sigma\rho^r_{12}(0)|\Biggr\}e^{-\Gamma_S \tau}\nonumber\\
&&+\widehat\gamma\,[\Sigma\rho_{11}(t_r)+2\Delta T{\rm Im}\Sigma\rho_{12}(t_r)]
e^{-\Gamma_L \tau}\nonumber\\
&&+{2|\epsilon|\over\cos\delta\phi}|\Sigma\rho^r_{12}(0)| e^{-\Gamma\tau}
\cos(\Delta m\tau-\phi-\delta\phi-\Sigma\phi_{12})\ ,\label{Sigmarho22^r}
\end{eqnarray}
where we have defined the phases $\Delta\phi_{12}$ and $\Sigma\phi_{12}$
through
\begin{eqnarray}
\Delta\rho^r_{12}(0)&=&
\left|\Delta\rho_{12}(t_r)-i\Delta
T[\Delta\rho_{11}(t_r)-\Delta\rho_{22}(t_r)]\right|e^{i\Delta\phi_{12}}\ ,\\
\Sigma\rho^r_{12}(0)&=&
\left|\Sigma\rho_{12}(t_r)-i\Delta
T[\Sigma\rho_{11}(t_r)-\Sigma\rho_{22}(t_r)]\right|e^{i\Sigma\phi_{12}}\ .
\end{eqnarray}
In these expressions, the ``initial-condition" input matrices $\Delta\rho(t_r)$
and $\Sigma\rho(t_r)$ are obtained from
Eqns.~(\ref{Deltarho11^0+1})--(\ref{Sigmarho12^0+1}) by inserting $t=t_r$. We
obtain a rather complicated result, which, in addition to the CPT-violating
parameters, also depends on $\Delta T$ and $t_r$. To illustrate the behavior of
$A^r_{2\pi}$ let us consider two limiting cases: $t_r\ll\tau_S$ and
$t_r\gg\tau_S$. For a regenerator very close to the origin ($t_r\ll\tau_S$) we
basically have $\Delta\rho(t_r)\approx\Delta\rho(0)$ and
$\Sigma\rho(t_r)\approx\Sigma\rho(0)$, as in Eq.~(\ref{InitialConditions}),
and we obtain
\begin{eqnarray}
\Delta\rho^r_{22}(\tau)&\to&
2|\epsilon|\,{\cos(\phi+\delta\phi)\over\cos\delta\phi}\,e^{-\Gamma_S \tau}
-{2|\epsilon|\over\cos\delta\phi}e^{-\Gamma\tau}
\cos(\Delta m\tau-\phi-\delta\phi )\ ,\\
\Sigma\rho^r_{22}(\tau)&\to&
(1-\widehat\gamma)e^{-\Gamma_S \tau}+\widehat\gamma e^{-\Gamma_L \tau}\ .
\end{eqnarray}
Neglecting $\widehat\gamma$ we find
\begin{equation}
A^r_{2\pi}(\tau)\to{2|\epsilon|\over\cos\delta\phi}
\left\{\cos(\phi+\delta\phi)-e^{{1\over2}(\Gamma_S-\Gamma_L)\tau}
\cos(\Delta m\tau-\phi-\delta\phi)\right\}
\end{equation}
Thus, when the regenerator is placed near the production point the effects of
$\Delta T$ drop out, and the result without a regenerator is recovered (see
Eq.~(\ref{CPTx}) dropping $\widehat\gamma$ and all second-order terms).

\begin{figure}[p]
\vspace{6in}
\includegraphics{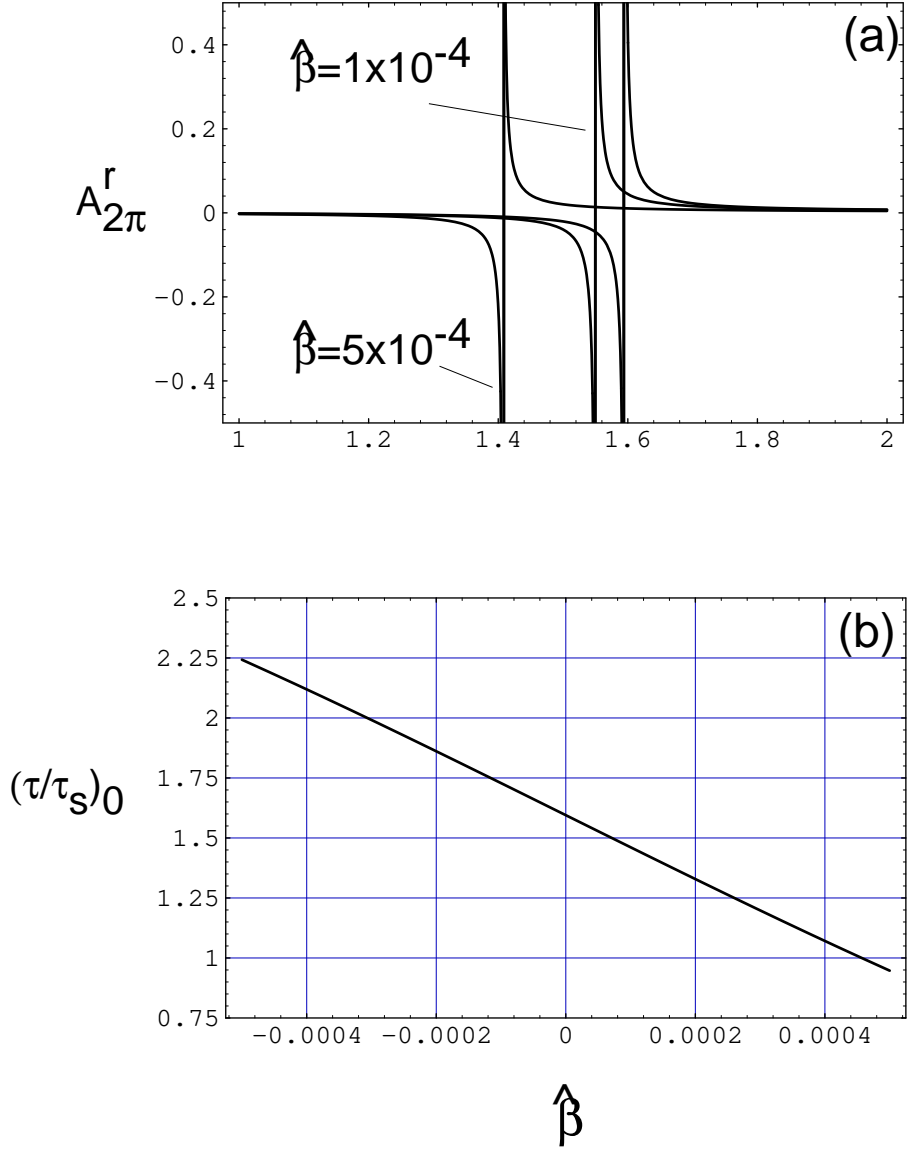}
\vspace{1.5cm}
\caption{The time-dependent asymmetry $A^r_{2\pi}(\tau)$ in the presence of a
thin regenerator placed far from the production point, as a function of the
time $\tau$ after leaving the regenerator, for representative choices of
$\widehat\beta$ ($A^r_{2\pi}(\tau)$ is rather insensitive to
$\widehat\alpha,\widehat\gamma$, which are set to zero). The right-most
curve corresponds to the standard case. Also shown (b) is the position of
the (first) zero in $A^r_{2\pi}(\tau)$ as a function of $\widehat\beta$.}
\label{A2pir-fig}
\end{figure}

Of more interest is the case of a regenerator placed in the asymptotic
region ($t_r\gg\tau_S$). In this case the expressions for $\Delta\rho(t_r)$
and $\Sigma\rho(t_r)$ simplify considerably, through first order:
\begin{equation}
\begin{tabular}{ll}
$\Delta\rho_{11}(t_r)\to{2|\epsilon|\cos(\phi-\delta\phi)\over\cos\delta\phi}$
&$\Sigma\rho_{11}(t_r)\to1+\widehat\gamma\approx1$\\
$\Delta\rho_{22}(t_r)\to0$&$\Sigma\rho_{22}(t_r)\to\widehat\gamma\approx0$\\
$\Delta\rho_{12}(t_r)\to0$&
$\Sigma\rho_{12}(t_r)\to{|\epsilon|\over\cos\delta\phi}e^{i(\delta\phi-\phi)}$
\end{tabular}
\end{equation}
Inserting these limiting expressions (and taking $\widehat\gamma=0$) we obtain
\begin{eqnarray}
\Delta\rho^r_{22}(\tau)&\to&
-2|\epsilon|\,{\cos(\phi+\delta\phi+\Delta\phi_{12})\over\cos\delta\phi}\,
|\Delta\rho^r_{12}(0)|e^{-\Gamma_S \tau}\nonumber\\
&&+{2|\epsilon|\over\cos\delta\phi}|\Delta\rho^r_{12}(0)|e^{-\Gamma\tau}
\cos(\Delta m\tau-\phi-\delta\phi-\Delta\phi_{12})\ ,\\
\Sigma\rho^r_{22}(\tau)&\to&
\Biggl\{2\Delta T|\epsilon|{\sin(\phi-\delta\phi)\over\cos\delta\phi}
-2|\epsilon|\,{\cos(\phi+\delta\phi+\Sigma\phi_{12})\over\cos\delta\phi}\,
|\Sigma\rho^r_{12}(0)|\Biggr\}e^{-\Gamma_S \tau}\nonumber\\
&&+{2|\epsilon|\over\cos\delta\phi}|\Sigma\rho^r_{12}(0)| e^{-\Gamma\tau}
\cos(\Delta m\tau-\phi-\delta\phi-\Sigma\phi_{12})\ ,
\end{eqnarray}
and thus
\begin{eqnarray}
A^r_{2\pi}(\tau)&\to&{|\Delta\rho^r_{12}(0)|\over|\Sigma\rho^r_{12}(0)|}
\left\{-\cos(\phi+\delta\phi+\Delta\phi_{12})
+e^{{1\over2}(\Gamma_S-\Gamma_L)\tau}
\cos(\Delta m\tau-\phi-\delta\phi-\Delta\phi_{12})\right\}\nonumber\\
&&/\Biggl\{
\left[{\Delta T\over|\Sigma\rho^r_{12}(0)|}\sin(\phi-\delta\phi)
-\cos(\phi+\delta\phi+\Sigma\phi_{12})\right]\nonumber\\
&&\qquad\qquad\qquad\qquad+e^{{1\over2}(\Gamma_S-\Gamma_L)\tau}
\cos(\Delta m\tau-\phi-\delta\phi-\Sigma\phi_{12})\Biggr\}
\label{A2pir}
\end{eqnarray}
with
\begin{eqnarray}
\Delta\rho^r_{12}(0)&\to&
\left|-i\Delta T{2|\epsilon|\cos(\phi-\delta\phi)\over\cos\delta\phi}
\right|e^{i\Delta\phi_{12}}\Rightarrow\Delta\phi_{12}=-\coeff{\pi}{2}\ ,\\
\Sigma\rho^r_{12}(0)&\to&
\left|{|\epsilon|\over\cos\delta\phi}e^{i(\delta\phi-\phi)}
-i\Delta T\right|e^{i\Sigma\phi_{12}}\ .
\end{eqnarray}
The result in Eq.~(\ref{A2pir}) reveals a large shift
($\Delta\phi_{12}=-\coeff{\pi}{2}$) in the interference pattern relative to the
case of no regenerator. According to our estimate of $\Delta T$ in
Eq.~(\ref{estimate}), it would appear that $\Delta T\gg|\epsilon|$ is a case
of interest to consider. In this limit, $\Delta T$ drops out from the
$A^r_{2\pi}$ observable, $\Delta\phi_{12}=\Sigma\phi_{12}=-\coeff{\pi}{2}$, and
\begin{equation}
A^r_{2\pi}(\tau)\to 2|\epsilon|{\cos(\phi-\delta\phi)\over\cos\delta\phi}\,
{\sin(\phi+\delta\phi)
+e^{{1\over2}(\Gamma_S-\Gamma_L)\tau}\sin(\Delta m\tau-\phi-\delta\phi)
\over
\sin(\phi+\delta\phi)-\sin(\phi-\delta\phi)
+e^{{1\over2}(\Gamma_S-\Gamma_L)\tau}\sin(\Delta m\tau-\phi-\delta\phi)}
\end{equation}
The time-dependence of $A^r_{2\pi}(\tau)$ is shown in Fig.~\ref{A2pir-fig} from
which it is apparent that $A^r_{2\pi}(\tau)$ is basically flat except for
values of $\tau$ for which $\sin(\Delta m\tau-\phi-\delta\phi)=0$. This occurs
for $(\tau/\tau_S)_0\approx2(n\pi+{\pi\over4}+\delta\phi)$, a result which
is plotted against $\widehat\beta$ (for $n=0$) also in Fig.~\ref{A2pir-fig}. We
note that for increasingly larger values of $n$, the structure in the curves
becomes narrower and therefore much less sensitive to $\widehat\beta$, with the
first zero ($n=0$) possibly being the only observable one.

\subsubsection{$R_{2\pi}$}
\label{sec:R2pi}
The observable $R_{2\pi}=R(K\to2\pi)$ has traditionally been the focus of
CP violation studies. Because the detector is physically located a distance
away from the source of the neutral kaons, most of the $K_S$ component of
the beam decays away, and one is basically sensitive only to the $K_L\to 2\pi$
decays. To study also the interesting interference region, a regenerator
is inserted in the path of the $K_L$ particles right before they reach the
detector, so that $K_S$ particles are regenerated and interference studies
are possible. Unfortunately, the regenerator complicates the physics somewhat.
To simplify the problem, let us first consider the case of a pure $K^0$
beam whose decay products can be detected from the instant of production
(not unlike in the CPLEAR experiment). We will address the effect of the
regenerator in the next subsection.

In our formalism, the $R_{2\pi}$ observable corresponds to the operator
${\cal O}_{2\pi}$ in (\ref{2pi-obs}), which gives
\begin{equation}
R_{2\pi}(t)=\rho_{22}(t)\ .
\label{R2pi}
\end{equation}
Through second order, the corresponding expression is obtained from
Eqs.~(\ref{rho22^0},\ref{rho22^1},\ref{rho22^2}) by inserting
$\rho_{11}(0)=\rho_{22}(0)=\rho_{12}(0)=1$. In the case of standard
quantum-mechanical CP violation, one obtains
\begin{equation}
R_{2\pi}(t)=c_S\, e^{-\Gamma_S t}+c_L\, e^{-\Gamma_L t}
+ 2c_I\, e^{-\Gamma t}\cos(\Delta mt-\phi)\ ,
\label{E.1}
\end{equation}
where to second order the $c_S,c_L,c_I$ coefficients are given by:
\begin{eqnarray}
c_S&=&1-2|\epsilon|\cos\phi+|\epsilon|^2(1+2\cos2\phi+t|\Delta\Gamma|)
\label{cSQM}\\
c_L&=&|\epsilon|^2
\label{cLQM}\\
c_I&=&|\epsilon|-2|\epsilon|^2\cos\phi
\label{cIQM}
\end{eqnarray}
It is then apparent that to the order calculated: $c_I^2=c_S c_L=|\epsilon|^2$.
Violations of this relation would indicate departures from standard quantum
mechanics, which can be parametrized by \cite{Eberhard}
\begin{equation}
\zeta=1-{c_I\over\sqrt{c_Sc_L}}\ .
\label{zeta}
\end{equation}
In our quantum-mechanical-violating framework we expect $\zeta\not=0$. Indeed,
we obtain
\begin{eqnarray}
c_S&=&1-\widehat\gamma-2|\epsilon|{\cos(\phi+\delta\phi)\over\cos\delta\phi}\\
c_L&=&\widehat\gamma+\widehat\gamma^2
+|\epsilon|^2{\cos(\phi-2\delta\phi)\over\cos\phi\cos^2\delta\phi}
-2|\epsilon|\widehat\gamma{\cos(\phi-\delta\phi)\over\cos\delta\phi}\\
c_I&=&{|\epsilon|\over\cos\delta\phi}
\end{eqnarray}
where only terms relevant to the computation of $\zeta$ to second order
have been kept (note that $\widehat\alpha$ does not contribute to $\zeta$ to
the order calculated). Also, in this case the general relation in
Eq.~(\ref{E.1}) gets modified by a phase shift in the interference term
$\phi\to\phi+\delta\phi$. Using these expressions we obtain\footnote{Note that
in the scenario discussed in Sec.~\ref{sec:A2pi}, where CPT violation accounts
for the observed CP violation (\ie, $|\epsilon|=0$,
$2\widehat\beta\cos\phi\to\pm|\epsilon|$, $\widehat\gamma\to2|\epsilon|^2$)
one obtains $c^2_I/(c_Sc_L)\to1\Leftrightarrow\zeta=0$. (This result was
implicitly obtained in Ref.~\cite{emncpt}.) Such result is not enough to
validate the scenario, since as discussed above, this scenario is fatally
flawed by the large phase shift in the interference term.}
\begin{equation}
{c^2_I\over c_S c_L}={{|\epsilon|^2/\cos^2\delta\phi}\over
\widehat\gamma(1-4|\epsilon|\cos\phi)
+|\epsilon|^2{\cos(\phi-2\delta\phi)\over\cos\phi\cos^2\delta\phi}}
\approx {1\over
{\widehat\gamma\over|\epsilon|^2}+{\cos(\phi-2\delta\phi)\over\cos\phi}}
\label{ratio}
\end{equation}
and thus
\begin{equation}
\zeta\approx{1\over2}\left[1-{1\over
{\widehat\gamma\over|\epsilon|^2}+{\cos(\phi-2\delta\phi)\over\cos\phi}}\right]
\approx
{\widehat\gamma\over2|\epsilon|^2}-{2\widehat\beta\over|\epsilon|}\sin\phi\ ,
\label{zeta-nor}
\end{equation}
where the second form holds for small values of $\widehat\gamma/|\epsilon|^2$
and $\delta\phi\approx-2\widehat\beta\cos\phi/|\epsilon|$. The parameter
$\zeta$ has been measured to be $\zeta^{\rm exp}=0.03\pm0.02$ \cite{Carithers}.
Setting $\widehat\beta=0$ one obtains
$\widehat\gamma\approx(3\pm2)\times10^{-7}$ \cite{Eberhard}. More generally,
the dependence of $\zeta$ on $\widehat\beta$ and $\widehat\gamma$ is shown
in Fig.~\ref{zeta-plot}, along with the present experimental limits on $\zeta$.

\subsubsection{$R^r_{2\pi}$}
Let us now turn to the $R^r_{2\pi}=\rho^r_{22}(\tau)$ observable in the
presence of a thin regenerator. Here $\rho^r_{22}(\tau)$ is given to first
order by Eqns.~(\ref{rho22^0},\ref{rho22^1}) with
$\rho(0)\to\rho^r(0)=\rho(t_r)+\delta\rho(t_r)$, and $\delta\rho$ given in
Eqn.~(\ref{4.6.7}). We obtain
\begin{eqnarray}
R^r_{2\pi}(\tau)&=&
\Biggl\{[\rho_{22}(t_r)-2\Delta T{\rm Im}\rho_{12}(t_r)]-
\widehat\gamma\,[\rho_{11}(t_r)+2\Delta T{\rm Im}\rho_{12}(t_r)]
\nonumber\\
&&-2|\epsilon|\,{\cos(\phi+\delta\phi+\phi_{12})\over\cos\delta\phi}\,
|\rho^r_{12}(0)|\Biggr\}e^{-\Gamma_S \tau}\nonumber\\
&&+\widehat\gamma\,[\rho_{11}(t_r)+2\Delta T{\rm Im}\rho_{12}(t_r)]
e^{-\Gamma_L \tau}\nonumber\\
&&+{2|\epsilon|\over\cos\delta\phi}|\rho^r_{12}(0)|e^{-\Gamma\tau}
\cos(\Delta m\tau-\phi-\delta\phi-\phi_{12})\label{rho22^r}\ ,
\label{129}
\end{eqnarray}
where
\begin{equation}
\rho^r_{12}(0)=\left|\rho_{12}(t_r)-i\Delta
T[\rho_{11}(t_r)-\rho_{22}(t_r)]\right|e^{i\phi_{12}}\ .
\end{equation}
As we discussed above, the initial condition matrix $\rho(t_r)$ is simply
$\rho_L$, namely
\begin{equation}
\rho(t_r)=\left( \begin{array}{cc}
1+\widehat\gamma-2|\epsilon|\cos\phi+4\widehat\beta\sin\phi\cos\phi
 &(|\epsilon | + i 2 {\widehat \beta }\cos\phi )e^{i\phi}  \\
(|\epsilon | - i 2 {\widehat \beta }\cos\phi )e^{-i\phi}  &
|\epsilon |^2 + {\widehat \gamma} -
4{\widehat \beta}^2 \cos ^2\phi - 4 {\widehat \beta} |\epsilon |\sin \phi
\end{array} \right)\ .
\label{rhoLr}
\end{equation}
Note that at the instant the beam leaves the regenerator ($\tau=0$), we
obtain $R^r_{2\pi}(0)=\rho_{22}^r(0)=\rho_{22}(t_r)-2\Delta T{\rm
Im}\rho_{12}(t_r)$ which, after inserting $\rho(t_r)$ from Eq.~(\ref{rhoLr}),
agrees with the result derived above in Eq.~(\ref{4.6.15}) where no time
dependence after leaving the regenerator was considered.

In the interference region the expression for $R_{2\pi}$ simplifies
considerably: we keep only the term proportional to $e^{-\Gamma\tau}$,
\begin{equation}
R^{\rm int}_{2\pi}(\tau)=
{2|\epsilon|\over\cos\delta\phi}|\rho^r_{12}(0)|e^{-\Gamma\tau}
\cos(\Delta m\tau-\phi-\delta\phi-\phi_{12})\ ,
\label{R2pi-int}
\end{equation}
with
\begin{equation}
\rho^r_{12}(0)\approx\left|{|\epsilon|\over\cos\delta\phi}
e^{i(\delta\phi-\phi)}-i\Delta T\right|e^{i\phi_{12}}\ .
\end{equation}
In this case we again see that the regenerator introduces a shift in the
interference pattern and modifies its overall magnitude, even in the absence
of CPT violation. In the limit in which $\Delta T\gg|\epsilon|$,
$|\rho_{12}^r(0)|\to\Delta T$, $\phi_{12}\to-\coeff{\pi}{2}$ and
\begin{equation}
R_{2\pi}^{\rm int}(\tau)\to {2|\epsilon|\Delta T\over\cos\delta\phi}
e^{-\Gamma\tau}\cos(\Delta m\tau-\phi-\delta\phi+\coeff{\pi}{2})\ ,
\label{2pi-reg}
\end{equation}
which exhibits a large phase shift and a distinctive linear dependence on
$\Delta T$, which is a nice signature. Moreover, the result still allows
a determination of the CPT-violating parameter $\beta$, through $\delta\phi$
(\ref{deltaphi}).

We now address the $\zeta$ parameter in the presence of a regenerator. Let
us first start with the case of standard quantum mechanics, where we expect
$\zeta$ to vanish. Looking back at Eqs.~(\ref{cSQM},\ref{cLQM},\ref{cIQM}),
we see that (to the order calculated) the $\zeta=0$ relation amounts to
$[c^{(1)}_I]^2=c^{(0)}_S c^{(2)}_2$, where the orders at which the relevant
contributions appear have been indicated. In the case of a regenerator, the
time dependence of $\rho^r_{22}(\tau)$ is the same as that of $\rho_{22}(t)$,
the only difference being in the coefficients which depend on different
initial-condition matrices ($\rho^r(0)$ versus $\rho(0)$). To make our result
more general, we will keep this initial-condition matrix unspecified. Using
Eqns.~(\ref{rho22^0},\ref{rho22^1},\ref{rho22^2}) we then get
\begin{eqnarray}
c^{(0)}_S&=&\rho_{22}(0)\\
c^{(2)}_L&=&\rho_{11}(0)|\epsilon|^2\\
c^{(1)}_I&=&|\rho_{12}(0)||\epsilon|
\end{eqnarray}
and therefore
\begin{equation}
\zeta_{\rm QM}=1-{c_I\over\sqrt{c_Sc_L}}=
1-{|\rho_{12}(0)|\over\sqrt{\rho_{11}(0)\rho_{22}(0)}}=0\ ,
\label{zetaQM}
\end{equation}
where we have used the fact that a pure quantum-mechanical ($2\times2$)
density matrix has zero determinant ($\det\rho(0)=\rho_{11}(0)\rho_{22}(0)
-|\rho_{12}(0)|^2$). This result applies immediately to the regenerator
case where a particular form of $\rho(0)$ is used, namely:
$\rho^r_{11}(0)\approx1$,
$\rho^r_{22}(0)\approx|\epsilon|^2+2\Delta T|\epsilon|\sin\phi$,
$|\rho^r_{12}(0)|^2\approx|\epsilon|^2+2\Delta T|\epsilon|\sin\phi$,
which indeed satisfy $\det\rho^r(0)=0$.

We now repeat the exercise in our quantum-mechanics-violating framework,
where we obtain
\begin{eqnarray}
c^{(0+1)}_S&=&\rho_{22}(0)-\rho_{11}(0)\widehat\gamma-2|\epsilon||\rho_{12}(0)|
\,{\cos(\phi+\delta\phi+\phi_{12})\over\cos\delta\phi}\\
c^{(1+2)}_L&=&\rho_{11}(0)\widehat\gamma+\rho_{22}(0)\widehat\gamma^2
+\rho_{11}(0)|\epsilon|^2
\,{\cos(\phi-2\delta\phi)\over\cos\phi\cos^2\delta\phi}
-2|\epsilon|\widehat\gamma|\rho_{12}(0)|
\,{\cos(\phi-\delta\phi-\phi_{12})\over\cos\delta\phi}\nonumber\\
&&\\
c^{(1)}_I&=&{|\epsilon|\over\cos\delta\phi}|\rho_{12}(0)|
\end{eqnarray}
which entail
\begin{eqnarray}
{c^2_I\over c_S c_L}&=&{|\epsilon|^2\over\cos^2\delta\phi}|\rho_{12}(0)|^2
\nonumber\\
&&/\Biggl\{
\rho_{11}(0)\rho_{22}(0)\widehat\gamma
+[\rho^2_{22}(0)-\rho^2_{11}(0)]\widehat\gamma^2
+\rho_{11}(0)\rho_{22}(0)|\epsilon|^2
{\cos(\phi-2\delta\phi)\over\cos\phi\cos^2\delta\phi}\nonumber\\
&&-{2|\epsilon|\widehat\gamma\over\cos\delta\phi}|\rho_{12}(0)|
[\rho_{22}(0)\cos(\phi-\delta\phi-\phi_{12})
+\rho_{11}(0)\cos(\phi+\delta\phi+\phi_{12})]\Biggr\}\nonumber\\
&&
\label{zeta_r}
\end{eqnarray}
This expression can be most easily interpreted in the limit of interest,
$\Delta T\gg|\epsilon|$, where the initial condition matrix $\rho^r(0)$
reduces to
\begin{eqnarray}
\rho^r_{11}(0)&\approx&1\ ,\\
\rho^r_{22}(0)&\approx&\widehat\gamma
+2\Delta T|\epsilon|\,{\sin(\phi-\delta\phi)\over\cos\delta\phi}
+|\epsilon|^2\,{\cos(\phi-2\delta\phi)\over\cos\phi\cos^2\delta\phi}\ ,\\
|\rho^r_{12}(0)|^2&\approx&2\Delta
 T|\epsilon|\,{\sin(\phi-\delta\phi)\over\cos\delta\phi}
+{|\epsilon|^2\over\cos^2\delta\phi}\ .
\end{eqnarray}
Note that the source of quantum mechanical decoherence is given by
\begin{equation}
\det\rho^r(0)\approx\widehat\gamma-2|\epsilon|^2
\,{\sin(\phi-\delta\phi)\sin\delta\phi\over\cos\phi\cos^2\delta\phi}
 \approx\widehat\gamma\ .
\end{equation}
With these expressions for $\rho^r(0)$ one obtains for the numerator and
denominator of Eq.~(\ref{zeta_r})
\begin{eqnarray}
c^2_I&\approx&{|\epsilon|^2\over\cos^2\delta\phi}
\left[2\Delta T|\epsilon|\,{\sin(\phi-\delta\phi)\over\cos\delta\phi}\right]\\
c_S c_L&\approx& 2\Delta
T|\epsilon|\,{\sin(\phi-\delta\phi)\over\cos\delta\phi}
\left[\widehat\gamma
+|\epsilon|^2\,{\cos(\phi-2\delta\phi)\over\cos\phi\cos^2\delta\phi}\right]
\end{eqnarray}
and thus the regenerator effects ($\Delta T$) drop out, and the expressions
without a regenerator in Eqs.~(\ref{ratio},\ref{zeta-nor}) are recovered, \ie,
$\zeta_r=\zeta$. This result also implies that the experimental limits on
$\zeta$ that are derived in the presence of a regenerator can be directly
applied to our expression for $\zeta$, as assumed in the previous subsection.

We note that, although the study of $\zeta$ alone, in tests using a regenerator
\cite{Eberhard}, does not seem to add anything to the discussion of
the possible breakdown of quantum-mechanical coherence within our framework,
individual terms in the expression (\ref{129}) for $R_{2\pi}^r(\tau)$ depend
linearly on the regenerator density via $\Delta T$, and the dependence on the
non-quantum-mechanical parameters is different from the no-regenerator case, so
the regenerator is able to provide interesting new probes of our framework.
In this respect, experimental tests of CPT-symmetry within quantum mechanics
suggested earlier~\cite{briere}, using arrays of regenerators, find also a
natural application within our quantum-mechanics-violating framework.

\subsubsection{$A^r_{\rm CPT}$}
In Sec.~\ref{sec:A_CPT} we showed that there is no contribution to the
$A_{\rm CPT}$ observable up to second order. One may wonder whether the
introduction of a regenerator could change this result. To this end we
compute $A_{\rm CPT}^r$, which is defined as in Eq.~(\ref{Acptf}) but with
the $\Delta\rho,\Sigma\rho$ matrices replaced by the
$\Delta\rho^r,\Sigma\rho^r$ matrices. Expressions for the latter are
complicated, as exhibited explicitly in the previous subsections. However, the
expression for $A_{\rm CPT}^r$ simplifies considerably when calculated
consistently through first order only, since many of the entries in the input
matrices $\Delta\rho(t_r),\Sigma\rho(t_r)$ need to be evaluated only to zeroth
order. After some algebra we obtain
\begin{equation}
A_{\rm CPT}^r(\tau)=2\Delta T
{[e^{-\Gamma t_r}\sin(\Delta m t_r)](e^{-\Gamma_L\tau}-e^{-\Gamma_S\tau})
+[e^{-\Gamma_L t_r}-e^{-\Gamma_S t_r}] e^{-\Gamma\tau}\sin\Delta (m\tau)
\over
e^{-\Gamma_L t_r}\,e^{-\Gamma_L \tau}+e^{-\Gamma_S t_r}\,e^{-\Gamma_S \tau}
+2e^{-\Gamma t_r}\,e^{-\Gamma\tau}\cos(\Delta m \tau+\Delta m t_r)}\ ,
\end{equation}
which for $\tau\gg\tau_S$ asymptotes to
\begin{equation}
A_{\rm CPT}^r(\tau)\to 2\Delta T\, e^{[-{1\over2}(\Gamma_S-\Gamma_L)t_r]}
\sin(\Delta mt_r)\ .
\end{equation}
Thus we see that all dependence on the CP- ($|\epsilon|$)
and CPT- ($\alpha,\beta,\gamma$) violating parameters drops out, which confirms
the result obtained without a regenerator. The novelty is that $A_{\rm CPT}^r$
is nonetheless non-zero, and proportional to $\Delta T$. This result is
interesting, but not unexpected since the matter in the regenerator scatters
$K^0$ differently from $\bar K^0$ (\ref{4.6.2}). Formally, this is expressed by
the fact that the regenerator Hamiltonian in Eq.~(\ref{4.6.3}) is proportional
to $\sigma_1$, therefore does not commute with the CPT operator, and so
violates CPT. That is, the regenerator is a CPT-violating environment, although
completely within standard quantum mechanics.

\section{Indicative Bounds on {\hbox{CPT-Violating~Parameters}}}
\label{sec:Bounds}
The formulae derived above are ready to be used in fits to the experimental
data. A complete analysis requires a detailed understanding of all the
statistical and systematic errors, and their correlations, which goes beyond
the scope of this paper \cite{elmncplear}. Here we restrict ourselves to
indications of the magnitudes of the bounds that are likely to be obtained from
such an analysis.

The parameter $\widehat\alpha$ can be constrained by observing that the overall
size of the interference term in $A_{2\pi}$ (\ref{CPTx}) does not differ
significantly from the standard result [see also Fig.~\ref{A2pi}(a)]. The
relevant dependence on $\widehat\alpha$ comes at second order through $\Delta
X_3$, which is given in Eq.~(\ref{DeltaX3}). From this expression we can see
that the dominant term is the third one, \ie,
$(-2|\epsilon|\widehat\alpha/\cos\delta\phi)t|\Delta\Gamma|\cos(\Delta
mt-\phi-\delta\phi)$, which is enhanced relative to the other terms because
of the $t|\Delta\Gamma|$ factor. The dominant interference term
through second order is then
\begin{equation}
-{2|\epsilon|\over\cos\delta\phi}\,[1+\widehat\alpha t|\Delta\Gamma|]\,
e^{{1\over2}(\Gamma_S-\Gamma_L)t}\cos(\Delta mt-\phi-\delta\phi)\ .
\end{equation}
For our indicative purposes, we assume that the size of the interference term
is within 5\% of the standard result for observations in the range
$t/\tau_S\lsim10$. Since $\widehat\alpha>0$ and the overall factor
$(1/\cos\delta\phi)\approx1$ (see below), we require
$\widehat\alpha t|\Delta\Gamma|\lsim0.05$ \cite{Lopez}, \ie,
\begin{equation}
\widehat\alpha\lsim5.0\times10^{-3}\ ,\qquad
\alpha\lsim3.7\times10^{-17}\GeV\ .
\label{boundalpha}
\end{equation}
This is to be compared to the order of magnitude
${\cal O}((\Lambda _{\rm QCD}~{\rm or}~m_s )^2 /M_{Pl} ) \lsim 10 ^{-19} \GeV $
which is of theoretical interest in the neutral kaon system.

The simplest way to constrain the parameter ${\widehat \beta }$
involves the observables $R_{2\pi}$ and $A_{2\pi}$, which differ
from the standard results at first order in $\widehat\beta$, as seen in
Fig.~\ref{A2pi}(b). This new contribution can affect the overall size of the
interference pattern and shift its phase relative to the superweak phase $\phi
$, as seen in equations (\ref{CPTx}) and (\ref{R2pi-int}). It is easy to check
that the shift in phase $\delta\phi$ is sufficiently small for any possible
change in the overall size of the interference pattern (due to $\delta\phi$) to
be negligible, \eg, $|\delta \phi | < 2^\circ $ implies a change
in the size by $ < 6 \times 10^{-4} $. There are two independent
sets of data that give information on $\delta \phi $ :
(i) the Particle Data Group compilation \cite{pdg} which fits
NA31, E731 and earlier data, and (ii) more recent data
from the E773 Collaboration \cite{E773,E773-prl}. New data from the
CPLEAR collaboration are discussed elsewhere \cite{elmncplear}.
In each case, both the superweak phase $\phi $ and the
$K \rightarrow \pi ^+ \pi ^- $ interference phase $\phi _{+-} $
are measured, and the corresponding values of
$\delta \phi = \phi _{+-} - \phi $ are extracted :
\begin{equation}
 \delta \phi = (-0.71 \pm 0.95 )^\circ~\cite{pdg}, \qquad
\delta\phi=(-0.84 \pm 1.42 )^\circ~\cite{E773}\ .
\label{6.1}
\end{equation}
Combining these independent measurements in quadrature,
we find $\delta \phi = (- 0.75 \pm 0.79)^\circ $,
corresponding to
\begin{equation}
 \widehat\beta=(2.0\pm2.2)\times10^{-5}\ ,\qquad
\beta=(1.5\pm1.6)\times10^{-19}\GeV\ ,
\label{6.2}
\end{equation}
to be compared with the earlier bound
$|{\widehat \beta }| \lsim 6 \times 10^{-5} $
obtained in ref.~\cite{Lopez} by demanding
$|\delta \phi | \lsim 2^\circ $.  As expected from Fig.~\ref{A2pi}, the
indicative bound (\ref{6.2}) on $|{\widehat \beta }|$ is considerably
more restrictive than that (\ref{boundalpha}) on $|{\widehat \alpha }|$.
Alternatively, one may bound $\widehat\beta$ by considering the relationship
(see \eg, \cite{E773-prl})
\begin{equation}
|m_{K^0}-m_{\bar K^0}|\approx 2\Delta m
{|\eta_{+-}|\over\sin\Phi_{\rm sw}}\,|\Phi_{+-}-\Phi_{\rm
sw}+\coeff{1}{3}\Delta\Phi|\ ,
\end{equation}
where $\Delta\Phi=\Phi_{00}-\Phi_{+-}$. In our framework, up to
$\epsilon'/\epsilon$ effects, $\Delta\Phi=0$, $\Phi_{\rm sw}=\phi$,
$\Phi_{+-}=\phi+\delta\phi$,
$|\eta_{+-}|=|\epsilon|/\cos\delta\phi\approx|\epsilon|$, and thus
\begin{equation}
|m_{K^0}-m_{\bar K^0}|\approx  2\Delta m
{|\epsilon||\delta\phi|\over\sin\phi}\approx2|\beta|\ .
\label{MassDiff}
\end{equation}
The E773 Collaboration has determined that \cite{E773-prl}
$|m_{K^0}-m_{\bar K^0}|/m_{K^0}<7.5\times10^{-19}$ at the 90\% CL, and thus it
follows that $|\widehat\beta|<2.6\times10^{-5}$,
$|\beta|<1.9\times10^{-19}\GeV$. This result is consistent
with that in Eq.~(\ref{6.2}).

\begin{table}[t]
\begin{center}
\caption{Compilation of indicative bounds on CPT-violating parameters and
their source.}
\label{Table1}
\smallskip
\begin{tabular}{lcl}
\underline{Source}&\qquad\qquad&\underline{Indicative bound}\\
$R_{2\pi},A_{2\pi}$&&$\widehat\alpha<5.0\times10^{-3}$\\
$R_{2\pi},A_{2\pi}$&&$\widehat\beta=(2.0\pm2.2)\times10^{-5}$\\
$|m_{K^0}-m_{\bar K^0}|$&&$\widehat\beta<2.6\times10^{-5}$\\
$R_{2\pi}$&&$\widehat\gamma\lsim5\times10^{-7}$\\
$\zeta$&&
${\widehat\gamma\over2|\epsilon|^2}-{2\widehat\beta\over|\epsilon|}\sin\phi
=0.03\pm0.02$\\
Positivity&&$\widehat\alpha>\widehat\beta^2/\widehat\gamma_{\rm max}
\sim(10^3\widehat\beta)^2$
\end{tabular}
\end{center}
\hrule
\end{table}

The $\widehat\gamma$ parameter has the peculiar property of appearing
in the observables at first order, but without being accompanied by a
similar first-order term proportional to $|\epsilon|$ (as is the case for
$\widehat\beta$). In fact, if corresponding terms exist, they are proportional
to $|\epsilon|^2$. This means that large deviations from the usual results
would occur unless $\widehat\gamma\lsim|\epsilon|^2$. This result is
exemplified in Fig.~\ref{A2pi}(c), from which we conclude that
$\widehat\gamma<10^{-5}$. In Ref.~\cite{Lopez}
$\widehat\gamma\lsim0.1|\epsilon|^2$ was obtained. However, since
$|\epsilon'/\epsilon|\sim10^{-3}$ effects have been neglected,
we conclude conservatively that
\begin{equation}
\widehat\gamma\lsim\left|{\epsilon'\over\epsilon}\right||\epsilon|\sim10^{-6}
\,,\qquad\gamma\lsim7\times10^{-20}\GeV
\ .
\label{gammabound}
\end{equation}

\begin{figure}[p]
\vspace{5in}
\includegraphics{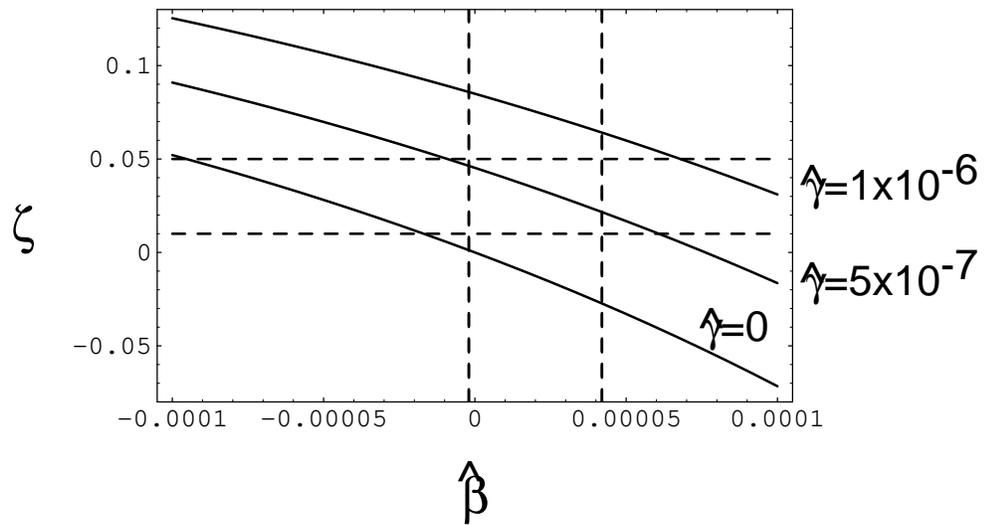}
\vspace{1.5cm}
\caption{The dependence of the quantum-mechanical-violating parameter
$\zeta$ on $\widehat\beta$ for representative values of $\widehat\gamma$
($\widehat\alpha$ does not contribute to the order calculated). The present
experimental value of $\zeta=0.03\pm0.02$ is indicated, as well as our
derived indicative bounds on $\widehat\beta=(2.0\pm2.2)\times10^{-5}$.}
\label{zeta-plot}
\end{figure}

We can also study the combined effects of $\widehat\beta$ and $\widehat\gamma$
on the $\zeta$ parameter in Eq.~(\ref{zeta-nor}), which reads
\begin{equation}
{\widehat\gamma\over2|\epsilon|^2}-{2\widehat\beta\over|\epsilon|}\sin\phi
=0.03\pm0.02\ .
\label{combined}
\end{equation}
The combined bounds on both parameters can be read off Fig.~\ref{zeta-plot},
which makes clearly the point that a combined fit is {\em essential} to obtain
the true bounds on the CPT-violating parameters. Note that the bounds on
$\widehat\beta$ (\ref{6.2}) and $\widehat\gamma$ (\ref{gammabound}) derived
above are consistent with those that follow from Eq.~(\ref{combined})
(see Fig.~\ref{zeta-plot}).

Let us close this section with a remark concerning the positivity constraints
in Eq.~(\ref{positivity}): $\alpha>0,\gamma>0$, and $\alpha\gamma>\beta^2$.
The data are not
yet sufficient to conclude anything about the sign of the
$\alpha$ and $\gamma$ parameters. The third constraint implies
\begin{equation}
\widehat\alpha>{\widehat\beta^2\over\widehat\gamma}>
{\widehat\beta^2\over\widehat\gamma_{\rm max}}\sim(10^3\,\widehat\beta)^2\ .
\end{equation}
Thus, if $\beta$ is observable, say $\widehat\beta\sim10^{-5}$, then
$\widehat\alpha>10^{-4}$ should be observable too.
A compilation of all these indicative bounds and their sources is given in
Table~\ref{Table1}.

\section{Comment on Two-Particle Decay Correlations}
\label{sec:HP}
Interesting further tests of quantum
mechanics and CPT symmetry can be devised by
exploiting initial-state correlations due to
the production of a pair of neutral kaons
in a pure quantum-mechanical state, \eg,
via $e^+e^- \rightarrow \phi \rightarrow K^0{\bar K}^0 $.
In this case, the initial state may
be represented by \cite{lipkin}
\begin{equation}
\ket{{\bf k}~;~{\bf - k}} = \coeff{1}{\sqrt{2}}
\left[ \ket{K^0 ({\bf k} )~;~{\bar K}^0 ({\bf - k})}
-\ket{{\bar K}^0 ({\bf k} )~;~K^0 ({\bf - k})} \right ]
\label{flavour}
\end{equation}
At subsequent times $t= t_1$ for particle $1$ and $t=t_2$ for particle $2$, the
joint probability amplitude is given in conventional quantum mechanics by
\begin{equation}
\ket{{\bf k},~t_1~;~{\bf - k},~t_2} \equiv
e^{-i H ({\bf k} ) t_1} e^{-i H ({\bf - k} ) t_2}
\ket{{\bf k}~;~{\bf - k}}
\label{lipkin}
\end{equation}
Thus the temporal evolution of the two-particle state is completely determined
by the one-particle variables (OPV) contained in $H$.

Tests of quantum mechanics and CPT symmetry in $\phi $ decays
have recently been discussed \cite{HP} in a conjectured
extension of the formalism of \cite{EHNS,emncpt}, in which
the density matrix of the two-particle system
was hypothesized to be described completely in terms
of such one-particle variables (OPV): namely $ H $ and
$(\alpha, \beta, \gamma )$. It was pointed out that this
OPV hypothesis had several striking consequences,
including apparent violations of energy conservation
and angular momentum.

As we have discussed above \cite{emnroma},
the only known theoretical framework in which
eq. (\ref{1.2}) has been derived is that of a non-critical
string approach to string theory, in which (i) energy is conserved
in the mean as a consequence of the renormalizability
of the world-sheet $\sigma$-model, but (ii) angular momentum
is not necessarily conserved \cite{emncpt,emnerice}, as this is not guaranteed
by renormalizability and is known to be violated
in some toy backgrounds \cite{emnroma},
though we cannot exclude the possibility that it may be conserved
in some particular backgrounds. Therefore, we are not concerned
that \cite{HP} find angular momentum non-conservation in their
hypothesized OPV approach. However, the absence of energy conservation
in their approach leads us to the conclusion
that irreducible two-particle parameters must be introduced
into the evolution of the two-particle
density matrix.  The appearance
of such non-local parameters does not concern
us, as the string is intrinsically non-local in target
space, and this fact plays a key role in
our model calculations of contributions to $\delta\H$.
The justification and parametrization of such irreducible
two-particle effects goes beyond the scope
of this paper, and we plan to study
this subject in more detail in due course.

\section{Conclusions}
\label{sec:conclusions}
We have derived in this paper approximate
expressions for a complete  set of neutral kaon
decay observables ($2\pi, 3\pi, \pi\ell\nu)$
which can be used to constrain parameters characterising CPT violation
in a formalism, motivated by ideas about quantum gravity
and string theory, that incorporates a possible microscopic
loss of quantum coherence by treating the neutral kaon
as an open quantum-mechanical system. Our explicit expressions
are to second order in the small CPT-violating parameters
$\alpha, \beta, \gamma $, and our systematic procedure for
constructing analytic approximations
may be extended to any desired level of accuracy.
Our formulae may be used to obtain indicative upper bounds
\begin{equation}
\alpha\lsim4\times10^{-17}\GeV\,,\quad
|\beta|\lsim3\times10^{-19}\GeV\,,\quad
\gamma\lsim7\times10^{-20}\GeV\,,
\label{8.1}
\end{equation}
which are comparable with the order of magnitude
$\sim 10^{-19} \GeV $ which theory indicates might be attained
by such CPT- and quantum-mechanics- violating parameters.
Detailed fits to recent CPLEAR experimental data
are reported elsewhere \cite{elmncplear}.

We have not presented explicit expressions for the case where
the deviation $|\epsilon '/\epsilon| \lsim 10^{-3} $ from pure
superweak CP violation is non-negligible, but our methods
can easily be extended to this case. They can also be used to obtain
more specific expressions for experiments with a regenerator,
if desired. The extension of the formalism of Ref. \cite{EHNS}
to correlated $K^0\bar K^0$ systems produced in $\phi $ decay, as at
DA$\Phi$NE \cite{dafnehb}, involves the introduction of two-particle variables,
which lies beyond the scope of this paper.

\begin{table}[t]
\begin{center}
\caption{Qualitative comparison of predictions for various observables
in CPT-violating theories beyond (QMV) and within (QM) quantum mechanics.
Predictions either differ ($\not=$) or agree ($=$) with the results obtained
in conventional quantum-mechanical CP violation. Note that these frameworks can
be qualitatively distinguished via their predictions for $A_{\rm T}$, $A_{\rm
CPT}$, $A_{\Delta m}$, and $\zeta$.}
\label{Table2}
\smallskip
\begin{tabular}{lcc}
\underline{Process}&QMV&QM\\
$A_{2\pi}$&$\not=$&$\not=$\\
$A_{3\pi}$&$\not=$&$\not=$\\
$A_{\rm T}$&$\not=$&$=$\\
$A_{\rm CPT}$&$=$&$\not=$\\
$A_{\Delta m}$&$\not=$&$=$\\
$\zeta$&$\not=$&$=$
\end{tabular}
\end{center}
\hrule
\end{table}

As mentioned in the main text, in Appendix~\ref{app:A} we have obtained
formulae for all observables in the case of CPT violation within standard
quantum mechanics. In the case of $A^{\rm QM}_{2\pi}$ and $A^{\rm QM}_{3\pi}$
one can ``mimic" the results from standard CP violation with suitable choices
of the CPT-violating parameters ($\delta M=0$, $\widehat{\delta\Gamma}\to
-2|\epsilon|/\cos\phi$). However, this possibility is experimentally excluded
because of the large value it entails for the $A_{\rm CPT}$ observable. In
passing we showed that the $\zeta$ parameter vanishes since no violations of
quantum mechanics are allowed. In analogy with Sec.~\ref{sec:Bounds},
we also obtained indicative bounds on the CPT-violating parameters.
In Table~\ref{Table2} we list all the observables and make a qualitative
comparison between them and conventional quantum-mechanical CP violation.
We see that the quantum-mechanical (QM) and quantum-mechanics-violating (QMV)
CPT-violating frameworks can be qualitatively distinguished by their
predictions for $A_{\rm T}$, $A_{\rm CPT}$, $A_{\Delta m}$, and $\zeta$.

We close by reiterating that the neutral kaon system
is the best microscopic laboratory for testing
quantum mechanics and CPT symmetry. We believe
that violations of these two fundamental principles,
if present at all, are likely to be linked, and have proposed
a formalism that can be used to explore systematically
this hypothesis, which is motivated by ideas about
quantum gravity and string theory. Our understanding
of these difficult issues is so incomplete
that we cannot calculate the sensitivity which would
be required to reveal modifications of quantum
mechanics or a violation of CPT. Hence we cannot
promise success in any experimental search for such
phenomena. However, we believe that both the theoretical
and experimental communities should be open to their
possible appearance.

\section*{Acknowledgments}
We would like to thank P. Eberhard, P. Huet, P. Pavlopoulos and T. Ruf for
useful discussions. The work of N.E.M. has been supported by a European Union
Research Fellowship, Proposal Nr. ERB4001GT922259, and that of D.V.N. has been
supported in part by DOE grant DE-FG05-91-ER-40633. J.E. thanks P. Sorba and
the LAPP Laboratory for hospitality during work on this subject. N.E.M. also
thanks D. Cocolicchio, G. Pancheri, N. Paver and other members of the
DA$\Phi$NE working groups for their interest in this work.

\newpage
\appendix

\section{CPT Violation in the Quantum-Mechanical Density Matrix
Formalism for Neutral Kaons}
\label{app:A}
In this appendix we review the density matrix formalism for neutral kaons and
CPT violation within the conventional quantum-mechanical
framework \cite{peccei,emncpt}. The time evolution of a generic density matrix
is determined in this case by the usual quantum Liouville equation
\begin{equation}
\partial _t \rho = -i (H\rho - \rho H^{\dagger})\ .
\label{denmatr}
\end{equation}
The conventional phenomenological Hamiltonian for the
neutral kaon system contains hermitian (mass) and antihermitian (decay)
components:
\begin{equation}
  H = \left( \begin{array}{cc}
 (M + \coeff{1}{2}\delta M) - \coeff{1}{2}i(\Gamma + \coeff{1}{2}
 \delta \Gamma)
&   M_{12}^{*} - \coeff{1}{2}i\Gamma _{12}^{*} \\
           M_{12}  - \coeff{1}{2}i\Gamma _{12}
&    (M - \coeff{1}{2}
    \delta M)-\coeff{1}{2}i(\Gamma
    - \coeff{1}{2}
    \delta \Gamma ) \end{array}\right)\ ,
\label{hmatr}
\end{equation}
in the ($K^0$, ${\bar K}^0$) basis. The ${\delta}M$ and ${\delta}{\Gamma}$
terms violate CPT \cite{peccei}. As in Ref.~\cite{EHNS},
we define components of $\rho$ and $H$ by
\begin{equation}
\rho \equiv  \coeff{1}{2}\rho _{\alpha} \sigma _{\alpha}
\qquad ; \qquad H \equiv \coeff{1}{2}h_{\alpha}\sigma _{\alpha}
\qquad ; \qquad \alpha = 0,1,2,3
\label{rhosigma}
\end{equation}
in a Pauli $\sigma$-matrix representation : the ${\rho_{\alpha}}$ are
real, but the $h_{\beta}$ are complex. The CPT transformation is
represented by
\begin{equation}
    {\rm CPT} \ket{K^{0}}=e^{i\theta}\ket{{\bar K}^{0}}, \qquad
{\rm CPT}  \ket{{\bar K}^{0}}=e^{-i\theta}\ket{K^{0}}  \ ,
\label{cpt}
\end{equation}
for some phase ${\theta}$, which is represented in our matrix formalism by
\begin{equation}
   {\rm CPT}  \equiv \left( \begin{array}{c}
 0   \qquad  e^{i\theta} \\
 e^{-i\theta}  \qquad 0 \end{array}\right)\ .
\label{cptmatr}
\end{equation}
Since this matrix is a linear combination of ${\sigma}_{1,2}$, CPT
invariance of the phenomenological Hamiltonian, $H$ =
$({\rm CPT})^{-1}H({\rm CPT})$, clearly requires that $H$ contain no term
proportional to ${\sigma}_3$, \ie, $h_3$ = $0$ so that ${\delta}M$ =
${\delta}{\Gamma}$ = $0$.

Conventional quantum-mechanical evolution is represented by
$\partial_t\rho_\alpha =H_{{\alpha}{\beta}}{\rho_{\beta}}$, where, in the
($K^0$, ${\bar K}^0$) basis and allowing for the possibility of CPT violation,
\begin{equation}
  H_{\alpha\beta} \equiv \left( \begin{array}{rrrr}
 {\rm Im}h_0 & {\rm Im}h_1 & {\rm Im}h_2 & {\rm Im}h_3 \\
 {\rm Im}h_1 & {\rm Im}h_0 & -{\rm Re}h_3 & {\rm Re}h_2 \\
 {\rm Im}h_2 & {\rm Re}h_3 & {\rm Im}h_0 & -{\rm Re}h_1 \\
 {\rm Im}h_3 & -{\rm Re}h_2 & {\rm Re}h_1 & {\rm Im}h_0 \end{array}\right)\ .
\label{habmatr}
\end{equation}
We note that the real parts of the matrix $h$ are
antisymmetric, whilst its imaginary parts are symmetric.
Now is an appropriate time to transform to the $K_{1,2}  =
\coeff{1}{\sqrt{2}}(K^0 \mp {\bar K}^0)$
basis, corresponding to $ \sigma_1\leftrightarrow\sigma_3$,
$\sigma_2\leftrightarrow-\sigma_2$, in which $H_{\alpha\beta }$ becomes
\begin{equation}
 H_{\alpha\beta}
 =\left( \begin{array}{cccc}  - \Gamma & -\coeff{1}{2}\delta \Gamma
& -{\rm Im} \Gamma _{12} & -{\rm Re}\Gamma _{12} \\
 - \coeff{1}{2}\delta \Gamma
  & -\Gamma & - 2{\rm Re}M_{12}&  -2{\rm Im} M_{12} \\
 - {\rm Im} \Gamma_{12} &  2{\rm Re}M_{12} & -\Gamma & -\delta M    \\
 -{\rm Re}\Gamma _{12} & -2{\rm Im} M_{12} & \delta M   & -\Gamma
\end{array}\right)\ .
\label{hcomp}
\end{equation}
The corresponding equations of motion for the components of $\rho$ in
the $K_{1,2}$ basis are [as above we neglect ${\rm Im}\Gamma_{12}$
contributions]
\begin{eqnarray}
\dot\rho_{11}&=&-\Gamma_L\rho_{11}
-2{\rm Re}\,[({\rm Im}M_{12}+\coeff{1}{4}\delta\Gamma+\coeff{i}{2}\delta
M)\rho_{12}]\,,\\
\dot\rho_{12}&=&-(\Gamma+i\Delta m)\rho_{12}
  +({\rm Im}M_{12}-\coeff{1}{4}\delta\Gamma-\coeff{i}{2}\delta M)\rho_{11}
-({\rm Im}M_{12}+\coeff{1}{4}\delta\Gamma-\coeff{i}{2}\delta M)\rho_{22},
\nonumber\\
&&\\
\dot\rho_{22}&=&-\Gamma_S\rho_{22}
+2{\rm Re}\,[({\rm Im}M_{12}-\coeff{1}{4}\delta\Gamma+\coeff{i}{2}\delta
M)\rho_{12}]\ .
\end{eqnarray}
One can readily verify that ${\rho}$ decays at large $t$ to
\begin{equation}
  \rho \sim e^{-\Gamma _Lt}
 \left( \begin{array}{cc}
 1   &  \epsilon^* + \delta^* \\
 \epsilon + \delta & |\epsilon + \delta |^2 \end{array}\right)\ ,
\label{rhodec}
\end{equation}
which has a vanishing determinant, thus corresponding to a pure long-lived mass
eigenstate $K_L$. The CP-violating parameter ${\epsilon}$ and the CPT-violating
parameter $\delta$ are given as above, namely
\begin{equation}
  \epsilon =\frac{{\rm Im} M_{12}}{\coeff{1}{2}|\Delta\Gamma|+i\Delta m }\
,\qquad
\delta= -\coeff{1}{2}\frac{ \coeff{1}{2}\delta \Gamma+i\delta M}
{\coeff{1}{2}|\Delta\Gamma|+i\Delta m}\ .
\label{eps-delta}
\end{equation}
Conversely, in the short-$t$ limit a $K_S$ state is represented by
\begin{equation}
  \rho \sim e^{-\Gamma _St}
 \left( \begin{array}{cc}
 |\epsilon - \delta |^2    &  \epsilon - \delta \\
 \epsilon^* - \delta^* & 1  \end{array}\right)\ ,
\label{rhodecs}
\end{equation}
which also has zero determinant. Note that the relative signs of the $\delta$
terms have reversed: this is the signature of CPT violation in the conventional
quantum-mechanical formalism. Note that the density matrices
(\ref{rhodec},\ref{rhodecs}) correspond to the state vectors
\begin{eqnarray}
\ket{K_L} &\propto& (1+\epsilon-\delta)\ket{K^0} -
 (1-\epsilon+\delta)\ket{{\bar K}^0}\\
\ket{K_S} &\propto& (1+\epsilon+\delta)\ket{K^0} +
 (1-\epsilon-\delta)\ket{{\bar K}^0}
\end{eqnarray}
and are both pure, as should be expected in conventional quantum mechanics,
even if CPT is violated.

As above, we solve the differential equations in perturbation theory in
$|\epsilon|$ and the new parameters
\begin{equation}
\widehat{\delta M}\equiv{\delta M\over|\Delta\Gamma|}\ ,\qquad
\widehat{\delta\Gamma}\equiv{\delta\Gamma\over|\Delta\Gamma|}\ .
\label{app.2}
\end{equation}
The zeroth order results for the $\rho_{ij}$
are the same as those in Eqs.~(\ref{rho11^0},\ref{rho22^0},\ref{rho12^0}),
namely
\begin{eqnarray}
\rho^{(0)}_{11}(t)&=&\rho_{11}(0)\,e^{-\Gamma_L t}\ ,\\
\rho^{(0)}_{22}(t)&=&\rho_{22}(0)\,e^{-\Gamma_S t}\ ,\\
\rho^{(0)}_{12}(t)&=&\rho_{12}(0)\,e^{-(\Gamma+i\Delta m t)}\ .
\end{eqnarray}
The first-order results for the density matrix elements are:
\begin{eqnarray}
\rho_{11}^{(1)}&=&-2|X'||\rho_{12}(0)|\left[
e^{-\Gamma_L t}\cos(\phi-\phi_{X'}-\phi_{12})
-e^{-\Gamma t}\cos(\Delta mt+\phi-\phi_{X'}-\phi_{12})\right]\\
\rho_{22}^{(1)}&=&-2|X||\rho_{12}(0)|\left[
e^{-\Gamma_S t}\cos(\phi+\phi_{X}+\phi_{12})
-e^{-\Gamma t}\cos(\Delta mt-\phi-\phi_{X}-\phi_{12})\right]\\
\rho_{12}^{(1)}&=&\rho_{11}(0)|X|e^{-i(\phi+\phi_X)}
\left[e^{-\Gamma_L t}-e^{-(\Gamma+i\Delta m)t}\right]
+\rho_{22}(0)|X'|e^{i(\phi-\phi_{X'})}
\left[e^{-\Gamma_S t}-e^{-(\Gamma+i\Delta m)t}\right]\nonumber\\
&&
\end{eqnarray}
where the two complex constants $X$ and $X'$ are defined by:
\begin{eqnarray}
X&=&|\epsilon|-\coeff{1}{2}\cos\phi\,\widehat{\delta\Gamma}
+i\cos\phi\,\widehat{\delta M}\, ,\qquad
\tan\phi_{X}={\cos\phi\,\widehat{\delta M}\over
|\epsilon|-\coeff{1}{2}\cos\phi\,\widehat{\delta\Gamma}}\ ,
\label{X}\\
X'&=&|\epsilon|+\coeff{1}{2}\cos\phi\,\widehat{\delta\Gamma}
+i\cos\phi\,\widehat{\delta M}\, ,\qquad
\tan\phi_{X'}={\cos\phi\,\widehat{\delta M}\over
|\epsilon|+\coeff{1}{2}\cos\phi\,\widehat{\delta\Gamma}}\ .
\label{X'}
\end{eqnarray}
For future reference, we note the special case that occurs when $\delta M=0$
and
$|\epsilon|=0$, namely
\begin{eqnarray}
\delta\Gamma>0:&&\phi_{X}=\pi,\quad \phi_{X'}=0\ ;\\
\delta\Gamma<0:&&\phi_{X}=0,\quad \phi_{X'}=\pi\ .
\end{eqnarray}

With the results for $\rho$ through first order, and inserting the appropriate
initial conditions (\ref{InitialConditions}), we can immediately write down the
expressions for the various observables discussed in Sec.~\ref{sec:analytical}.
For $A_{2\pi}$ we obtain
\begin{equation}
A^{\rm QM}_{2\pi}(t)={2|X|\cos(\phi+\phi_{X})
-2|X|e^{{1\over2}(\Gamma_S-\Gamma_L)t}\cos(\Delta mt-\phi-\phi_{X})\over
1+e^{(\Gamma_S-\Gamma_L)t}\,|X|^2}\ ,
\label{A2piQM}
\end{equation}
where in the denominator we have also included the non-negligible second-order
contributions to $\Sigma\rho_{22}^{(2)}$. From this expression it is
interesting to note that one can {\em mimic} the
standard CP-violating result for $A_{2\pi}$ in Eq.~(\ref{A2pi-usual}) by
setting $|\epsilon|\to0$ and making the following choices for the CPT-violating
parameters
\begin{equation}
{\rm mimic\ CP\ violation:}\qquad\qquad\delta M=0,\qquad
\widehat{\delta\Gamma}\to-{2|\epsilon|\over\cos\phi}\ ,
\label{mimic}
\end{equation}
which give $|X|\to|\epsilon|$ and $\phi_X=0$. For the $A_{3\pi}$ observable
we find
\begin{equation}
    A^{\rm QM}_{3\pi}(t) = 2|X'|\cos(\phi-\phi_{X'})
-2e^{-{1\over2}(\Gamma_S-\Gamma_L)t}
\left[{\rm Re}\eta_{3\pi}\cos\Delta mt - {\rm Im}\eta_{3\pi}\sin\Delta
 mt\right]\ ,
\end{equation}
with
\begin{equation}
{\rm Re}\eta_{3\pi}=|X'|\cos(\phi-\phi_{X'}),\quad
{\rm Im}\eta_{3\pi}=|X'|\sin(\phi-\phi_{X'})\ ,
\end{equation}
that is
\begin{equation}
{{\rm Im}\eta_{3\pi}\over{\rm Re}\eta_{3\pi}}=\tan(\phi-\phi_{X'})\ .
\end{equation}
Here we also note that the standard CP-violating result is obtained for
the choices of parameters in Eq.~(\ref{mimic}) which give $|X'|\to|\epsilon|$
and $\phi_{X'}=\pi$, since $\tan(\phi-\pi)=\tan\phi$.

For the observable $A_{\rm T}$, we obtain the following exactly
time-independent
first-order expression
\begin{equation}
A^{\rm QM}_{\rm T}=
2|X'|\cos(\phi-\phi_{X'})+2|X|\cos(\phi+\phi_X)=4|\epsilon|\cos\phi\ ,
\label{AQM_T}
\end{equation}
which is identical to the case of no CPT violation. In the case of $A_{\rm
CPT}$ we find
\begin{equation}
A^{\rm QM}_{\rm CPT}(t)={A_1(e^{-\Gamma_L t}-e^{-\Gamma_S t})
-2e^{-\Gamma t}A_2\sin\Delta m t
\over e^{-\Gamma_L t}+e^{-\Gamma_S t}-2 e^{-\Gamma t}\cos\Delta m t}\ ,
\label{AQM_CPT}
\end{equation}
with
\begin{eqnarray}
&&\!\!\!\!\!\!\!\!\!\!\!\!\!\!\!\!\!\!A_1=2|X'|\cos(\phi-\phi_{X'})
-2|X|\cos(\phi+\phi_{X})=
4\sin\phi\cos\phi\,\widehat{\delta M}+2\cos^2\phi\,\widehat{\delta\Gamma}\\
&&\!\!\!\!\!\!\!\!\!\!\!\!\!\!\!\!\!\!A_2=-2|X'|\sin(\phi-\phi_{X'})
+2|X|\sin(\phi+\phi_X)=
4\cos^2\phi\,\widehat{\delta M}-2\sin\phi\cos\phi\,\widehat{\delta\Gamma}
\end{eqnarray}
Note that $|\epsilon|$ drops out of the expression for $A_{\rm CPT}$ as it
should. In the long-time limit we obtain
\begin{equation}
A^{\rm QM}_{\rm CPT}\to4\sin\phi\cos\phi\,\widehat{\delta M}
+2\cos^2\phi\,\widehat{\delta\Gamma}\ .
\end{equation}
Since the dynamical equations determining the density matrix do not
manifestly possess the mimicking symmetry in Eq.~(\ref{mimic}), one expects
this mimicking phenomenon to break down in some observables. This is the
case of $A_{\rm CPT}$ where we find the following asymptotic
``mimic"  result
\begin{equation}
A_{\rm CPT}\to-4|\epsilon|\cos\phi\approx-6\times10^{-3}\ ,
\label{compare}
\end{equation}
to be contrasted with the standard result of $A_{\rm CPT}=0$. Experimentally,
the CPLEAR Collaboration has measured this parameter to be $A_{\rm CPT}^{\rm
exp}=(-0.4\pm2.0\pm2.0\pm1.5)\times10^{-3}$ \cite{Guyot}. Comparing
the prediction in Eq.~(\ref{compare}) with the experimental data, we see
that the ``mimic" result appears disfavored by the $A_{\rm CPT}$ measurement.

Finally, since
$\Delta\rho^{(1)}_{12}=\Sigma\rho^{(1)}_{11}=\Sigma\rho^{(1)}_{22}=0$, the
$A_{\Delta m}$ observable has the same first-order expression as in standard CP
violation, namely
\begin{equation}
A^{\rm QM}_{\Delta m}(t)=-{2e^{-\Gamma t}\cos\Delta m t\over e^{-\Gamma_L t}
+e^{-\Gamma_S t}}\ .
\end{equation}

Since in this mechanism of CPT violation quantum mechanics is not violated,
from the discussion in subsection~\ref{sec:R2pi} we expect the parameter
$\zeta$ to vanish. Indeed, using the above expressions for $\rho_{22}$ we find
\begin{eqnarray}
c^{(0)}_S&=&\rho_{22}(0)\\
c^{(2)}_L&=&\rho_{11}(0)|X|^2\\
c^{(1)}_I&=&|\rho_{12}(0)|\,|X|
\end{eqnarray}
where we have also calculated the needed second-order (long-lived) terms in
$\rho_{22}$. Moreover, the generic expression (\ref{E.1}) gets modified
in the interference term by the replacement: $\phi\to\phi+\phi_X+\phi_{12}$.
It then immediately follows that $c^2_I/(c_Sc_L)=|\rho_{12}(0)|^2/[\rho_{11}(0)
\rho_{22}(0)]=1$, where we have made use of the $\det\rho(0)=0$ property.
Therefore, as expected $\zeta=0$.

As in Sec.~\ref{sec:Bounds}, we can derive indicative bounds on the
CPT-violating parameters. The coefficient of the interference term in
$A_{2\pi}^{\rm QM}$ (\ref{A2piQM}) can be expressed as:
$|X|=||\epsilon|-{1\over2}\cos\phi\widehat{\delta\Gamma}|/\cos\phi_X$.
Demanding that this amplitude differ by less than 5\% from the usual case, and
with the a priori knowledge that $\phi_X$ should be small (as we demonstrate
below), we obtain ${1\over2}\cos\phi|\widehat{\delta\Gamma}|/|\epsilon|<0.05$,
\ie,
\begin{equation}
|\widehat{\delta\Gamma}|<3\times10^{-4}\,,\qquad
|\delta\Gamma|<2\times10^{-18}\GeV\ .
\label{deltaGammabound}
\end{equation}
We can obtain a bound on $\widehat{\delta M}$ by noticing the correspondence
$\delta M\leftrightarrow-2\beta$ that follows from
Eqs.~(\ref{deltaphi},\ref{X}) when the bound in Eq.~(\ref{deltaGammabound})
holds. From Eq.~(\ref{6.2}) we then find
\begin{equation}
\widehat{\delta M}=(-4.0\pm4.4)\times10^{-5}\,,\qquad
\delta M=(-3.0\pm3.2)\times10^{-19}\GeV\ .
\label{deltaMbound}
\end{equation}
Alternatively, the analogue of Eq.~(\ref{MassDiff}) is $|m_{K^0}-m_{\bar
K^0}|\approx|\delta M|$, which entails $|\delta M|<3.7\times10^{-19}\GeV$,
once the 90\%CL upper bound from E773 \cite{E773-prl} is inserted.

\section{Second-Order Contributions to the \hbox{Density Matrix}}
\label{app:B}
The second-order contributions to the density matrix in our
quantum-mechanical-violating framework can be obtained by using
Eq.~(\ref{solution}) with the first-order inputs $\rho^{(1)}_{11,22,12}$
given in Eqs.~(\ref{rho11^1},\ref{rho22^1},\ref{rho12^1}).\footnote{Expressions
for $\rho^{(2)}_{22,12}$ valid for a particular choice of initial conditions
were given in Ref.~\cite{Lopez}.} We obtain:
\begin{equation}
\rho_{11}^{(2)}=\sum_{k=1}^7 c^{[11]}_k R^{[11]}_k\,(t)\ ,
\label{rho11^2}
\end{equation}
where the time-dependent $R^{[11]}_k\,(t)$ functions are given by:
\begin{eqnarray}
R^{[11]}_1\,(t)&=&e^{-\Gamma_L t}-e^{-\Gamma_S t}\\
R^{[11]}_2\,(t)&=&t|\Delta\Gamma| e^{-\Gamma_L t}\\
R^{[11]}_3\,(t)&=&-e^{-\Gamma t}\cos(\Delta m t-\delta\phi-\phi_{12})
+e^{-\Gamma_L t}\cos(\delta\phi+\phi_{12})\\
R^{[11]}_4\,(t)&=&-e^{-\Gamma t}\sin(\Delta mt+\phi)+e^{-\Gamma_L t}\sin\phi\\
R^{[11]}_5\,(t)&=&-e^{-\Gamma t}\left[{|\Delta\Gamma|t\over2\cos\phi}
\cos(\Delta mt+\phi-\delta\phi-\phi_{12})
+\cos(\Delta mt+2\phi-\delta\phi-\phi_{12})\right]
\nonumber\\
&&+e^{-\Gamma_L t}\cos(2\phi-\delta\phi-\phi_{12})\\
R^{[11]}_6\,(t)&=&-e^{-\Gamma t}\cos(\Delta mt+2\phi-2\delta\phi)
+e^{-\Gamma_L t}\cos(2\phi-2\delta\phi)\\
R^{[11]}_7\,(t)&=&-e^{-\Gamma t}\cos(\Delta mt-2\delta\phi)
+e^{-\Gamma_L t}\cos(2\delta\phi)
\end{eqnarray}
and the $c^{[11]}_k$ coefficients are:
\begin{eqnarray}
c^{[11]}_1&=&-\rho_{11}(0)\widehat\gamma^2-\rho_{22}(0)|\epsilon|^2
{\cos(\phi+2\delta\phi)\over\cos\phi\cos^2\delta\phi}-2|\rho_{12}(0)||\epsilon|
\widehat\gamma\,{\cos(\phi+\delta\phi+\phi_{12})\over\cos\delta\phi}\nonumber\\
&&\\
c^{[11]}_2&=& \left[\widehat\gamma^2-|\epsilon|^2\,
{\cos(\phi-2\delta\phi)\over\cos\phi\cos^2\delta\phi}\right]\,\rho_{11}(0)\\
c^{[11]}_3&=&4|\epsilon|\widehat\gamma\,
{\cos\phi\over\cos\delta\phi}\,|\rho_{12}(0)|\\
c^{[11]}_4&=&-{4\widehat\alpha|\epsilon|\over\tan\phi}\,
{\cos(\delta\phi-\phi_{12})\over\cos\delta\phi}\,|\rho_{12}(0)|\\
c^{[11]}_5&=&4\widehat\alpha|\epsilon|\,{\cos\phi\over\cos\delta\phi}\,
|\rho_{12}(0)|\\
c^{[11]}_6&=&{2|\epsilon|^2\over\cos^2\delta\phi}\,\rho_{11}(0)\\
c^{[11]}_7&=&{2|\epsilon|^2\over\cos^2\delta\phi}\,\rho_{22}(0)
\end{eqnarray}
Analogously,
\begin{equation}
\rho_{22}^{(2)}=\sum_{k=1}^7 c^{[22]}_k R^{[22]}_k\,(t)\ ,
\label{rho22^2}
\end{equation}
where the time-dependent $R^{[22]}_k\,(t)$ functions are given by:
\begin{eqnarray}
R^{[22]}_1\,(t)&=&e^{-\Gamma_L t}-e^{-\Gamma_S t}\\
R^{[22]}_2\,(t)&=&t|\Delta\Gamma| e^{-\Gamma_S t}\\
R^{[22]}_3\,(t)&=&e^{-\Gamma t}\cos(\Delta m t-\delta\phi-\phi_{12})
-e^{-\Gamma_S t}\cos(\delta\phi+\phi_{12})\\
R^{[22]}_4\,(t)&=&e^{-\Gamma t}\sin(\Delta mt-\phi)+e^{-\Gamma_S t}\sin\phi\\
R^{[22]}_5\,(t)&=&e^{-\Gamma t}\left[{|\Delta\Gamma|t\over2\cos\phi}
\cos(\Delta mt-\phi-\delta\phi-\phi_{12})
-\cos(\Delta mt-2\phi-\delta\phi-\phi_{12})\right]
\nonumber\\
&&+e^{-\Gamma_S t}\cos(2\phi+\delta\phi+\phi_{12})\\
R^{[22]}_6\,(t)&=&-e^{-\Gamma t}\cos(\Delta mt-2\delta\phi)
+e^{-\Gamma_S t}\cos(2\delta\phi)\\
R^{[22]}_7\,(t)&=&-e^{-\Gamma t}\cos(\Delta mt-2\phi-2\delta\phi)
+e^{-\Gamma_S t}\cos(2\phi+2\delta\phi)
\end{eqnarray}
and the $c^{[22]}_k$ coefficients are:
\begin{eqnarray}
c^{[22]}_1&=&\rho_{22}(0)\widehat\gamma^2+\rho_{11}(0)|\epsilon|^2
{\cos(\phi-2\delta\phi)\over\cos\phi\cos^2\delta\phi}-2|\rho_{12}(0)||\epsilon|
\widehat\gamma\,{\cos(\phi-\delta\phi-\phi_{12})\over\cos\delta\phi}\nonumber\\
&&\\
c^{[22]}_2&=& \left[-\widehat\gamma^2+|\epsilon|^2\,
{\cos(\phi+2\delta\phi)\over\cos\phi\cos^2\delta\phi}\right]\,\rho_{22}(0)\\
c^{[22]}_3&=&4|\epsilon|\widehat\gamma\,
{\cos\phi\over\cos\delta\phi}\,|\rho_{12}(0)|\\
c^{[22]}_4&=&{4\widehat\alpha|\epsilon|\over\tan\phi}\,
{\cos(\delta\phi-\phi_{12})\over\cos\delta\phi}\,|\rho_{12}(0)|\\
c^{[22]}_5&=&-4\widehat\alpha|\epsilon|\,{\cos\phi\over\cos\delta\phi}\,
|\rho_{12}(0)|\\
c^{[22]}_6&=&{2|\epsilon|^2\over\cos^2\delta\phi}\,\rho_{11}(0)\\
c^{[22]}_7&=&{2|\epsilon|^2\over\cos^2\delta\phi}\,\rho_{22}(0)
\end{eqnarray}
Finally,
\begin{eqnarray}
\rho^{(2)}_{12}&=&{2\widehat\alpha\over\tan\phi}
\Biggl\{{2\widehat\alpha\over\tan\phi}|\rho_{12}(0)|\sin\phi_{12}\,
R^{[12]}_1\,(t,0)
-{4i\widehat\alpha\over\tan\phi}|\rho_{12}(0)|\, R^{[12]}_2\,(t)
\nonumber\\
&&-{|\epsilon|\over\cos\delta\phi}\,[\rho_{11}(0)R^{[12]}_1\,
(t,\phi-\delta\phi)
+\rho_{22}(0)R^{[12]}_1\,(t,-\phi-\delta\phi)]\nonumber\\
&&+{2i|\epsilon|\sin\phi\over\cos\delta\phi}
[\rho_{11}(0)\sin(\phi-\delta\phi)R^{[12]}_3\,(t)
+\rho_{22}(0)\sin(\phi+\delta\phi)R^{[12]}_4\,(t)]\Biggr\}\nonumber\\
&&+{|\epsilon|e^{i\delta\phi}\over\cos\delta\phi}\Biggl\{
\widehat\gamma[\rho_{22}(0)-\rho_{11}(0)]\,[R^{[12]}_3\,(t)+R^{[12]}_4\,(t)]
\nonumber\\
&&+{2|\epsilon|\over\cos\delta\phi}|\rho_{12}(0)|
[iR^{[12]}_1\,(t,-\delta\phi-\phi_{12})\nonumber\\
&&-\cos(\phi-\delta\phi-\phi_{12})R^{[12]}_3\,(t)
-\cos(\phi+\delta\phi+\phi_{12})R^{[12]}_4\,(t)]\Biggr\}
\end{eqnarray}
where the time-dependent functions $R^{[12]}_k\,(t)$ are given by
\begin{eqnarray}
R^{[12]}_1\,(t,a)&=&e^{-\Gamma t}\,
[e^{ia}\sin\Delta mt-(\Delta mt) e^{-i\Delta mt-ia}]\\
R^{[12]}_2\,(t)&=&\coeff{1}{4}e^{-\Gamma t}
\left\{e^{-i\phi_{12}}[\sin\Delta mt-(\Delta mt) e^{i\Delta mt}]
+i(\Delta mt)^2\, e^{-i\Delta mt+i\phi_{12}}\right\}\\
R^{[12]}_3\,(t)&=&e^{-i\phi}\,[e^{-\Gamma_L t}-e^{-(\Gamma+i\Delta m)t}]\\
R^{[12]}_4\,(t)&=&e^{i\phi}\,[e^{-\Gamma_S t}-e^{-(\Gamma+i\Delta m)t}]
\end{eqnarray}

\clearpage

\clearpage

\clearpage

\clearpage

\clearpage

\clearpage

\end{document}